
%
\def\unredoffs{}
\tolerance=1000\hfuzz=2pt
\catcode`\@=11 
\ifx\hyperdef\UNd@FiNeD\def\hyperdef#1#2#3#4{#4}\def\hyperref#1#2#3#4{#4}\def\href#1#2{#2}\fi

\magnification=1200\unredoffs\baselineskip=16pt plus 2pt minus 1pt
\def\Date#1{\vfill\leftline{#1}\tenpoint\supereject%
\footline={\hss\tenrm\hyperdef\hypernoname{page}\folio\folio\hss}}%

{\count255=\time\divide\count255 by 60 \xdef\hourmin{\number\count255}
 \multiply\count255 by-60\advance\count255 by\time
 \xdef\hourmin{\hourmin:\ifnum\count255<10 0\fi\the\count255}
}
\def\date{\number\day.\number\month.\number\year\ at \hourmin}


\def\nolabels{\def\wrlabeL##1{}\def\eqlabeL##1{}\def\reflabeL##1{}}
\def\writelabels{\def\wrlabeL##1{\leavevmode\vadjust{\rlap{\smash%
{\line{{\escapechar=` \hfill\rlap{\sevenrm\hskip.03in\string##1}}}}}}}%
\def\eqlabeL##1{{\escapechar-1\rlap{\sevenrm\hskip.05in\string##1}}}%
\def\reflabeL##1{\noexpand\llap{\noexpand\sevenrm\string\string\string##1}}}
\nolabels

\global\newcount\secno \global\secno=0
\global\newcount\meqno \global\meqno=1
\def\s@csym{}

\def\newsec#1\par{\global\advance\secno by1%
{\toks0{#1}\message{(\the\secno. \the\toks0)}}%
\global\subsecno=0\eqnres@t\let\s@csym\secsym\xdef\secn@m{\the\secno}\noindent
{\bf\hyperdef\hypernoname{section}{\the\secno}{\the\secno.} #1}%
\writetoca{{\string\hyperref{}{section}{\the\secno}{\bf \the\secno\quad}} {\bf #1}}\par%
\nobreak\medskip\nobreak\noindent\ignorespaces}
\def\eqnres@t{\xdef\secsym{\the\secno.}\global\meqno=1\bigbreak\bigskip}
\def\sequentialequations{\def\eqnres@t{\bigbreak}}\xdef\secsym{}

\global\newcount\subsecno \global\subsecno=0
\def\subsec#1\par{\global\advance\subsecno by1%
{\toks0{#1}\message{(\s@csym\the\subsecno. \the\toks0)}}%
\global\subsubsecno=0%
\ifnum\lastpenalty>9000\else\bigbreak\fi
\noindent{\it\hyperdef\hypernoname{subsection}{\secn@m.\the\subsecno}%
{\secn@m.\the\subsecno.} #1}\writetoca{\string\hskip1.45cm
{\string\hyperref{}{subsection}{\secn@m.\the\subsecno}{\secn@m.\the\subsecno.}}
{#1}}\par\nobreak\medskip\nobreak\noindent\ignorespaces}

\def\appendix#1#2{\global\meqno=1\global\subsecno=0\xdef\secsym{\hbox{#1.}}%
\bigbreak\bigskip\noindent{\bf Appendix \hyperdef\hypernoname{appendix}{#1}%
{#1.} #2}{\toks0{(#1. #2)}\message{\the\toks0}}%
\xdef\s@csym{#1.}\xdef\secn@m{#1}%
\writetoca{{\string\hyperref{}{appendix}{#1}{\bf {#1}\quad}} {\bf #2}}%
\par\nobreak\medskip\nobreak}

%
\def\checkm@de#1#2{\ifmmode{\def\f@rst##1{##1}\hyperdef\hypernoname{equation}%
{#1}{#2}}\else\hyperref{}{equation}{#1}{#2}\fi}
\def\eqnn#1{\DefWarn#1\xdef #1{(\noexpand\relax\noexpand\checkm@de%
{\s@csym\the\meqno}{\secsym\the\meqno})}%
\wrlabeL#1\writedef{#1\leftbracket#1}\global\advance\meqno by1}
\def\f@rst#1{\c@t#1a\em@ark}\def\c@t#1#2\em@ark{#1}
\def\eqna#1{\DefWarn#1\wrlabeL{#1$\{\}$}%
\xdef #1##1{(\noexpand\relax\noexpand\checkm@de%
{\s@csym\the\meqno\noexpand\f@rst{##1}1}{\hbox{$\secsym\the\meqno##1$}})}
\writedef{#1\numbersign1\leftbracket#1{\numbersign1}}\global\advance\meqno by1}
\def\eqn#1#2{\DefWarn#1%
\xdef #1{(\noexpand\hyperref{}{equation}{\s@csym\the\meqno}%
{\secsym\the\meqno})}$$#2\eqno(\hyperdef\hypernoname{equation}%
{\s@csym\the\meqno}{\secsym\the\meqno})\eqlabeL#1$$%
\writedef{#1\leftbracket#1}\global\advance\meqno by1}
\def\xeqn{\expandafter\xe@n}\def\xe@n(#1){#1}
\def\xeqna#1{\expandafter\xe@n#1}
\def\eqns#1{(\e@ns #1{\hbox{}})}
\def\e@ns#1{\ifx\UNd@FiNeD#1\message{eqnlabel \string#1 is undefined.}%
\xdef#1{(?.?)}\fi{\let\hyperref=\relax\xdef\next{#1}}%
\ifx\next\em@rk\def\next{}\else%
\ifx\next#1\xeqn#1\else\def\n@xt{#1}\ifx\n@xt\next#1\else\xeqna#1\fi
\fi\let\next=\e@ns\fi\next}

\def\DefWarn#1{\ifx\UNd@FiNeD#1\else
\immediate\write16{*** WARNING: the label \string#1 is already defined ***}\fi}
%
\newskip\footskip\footskip14pt plus 1pt minus 1pt 
\def\footnotefont{\ninepoint}\def\f@t#1{\footnotefont #1\@foot}
\def\f@@t{\baselineskip\footskip\bgroup\footnotefont\aftergroup\@foot\let\next}
\setbox\strutbox=\hbox{\vrule height9.5pt depth4.5pt width0pt}
\global\newcount\ftno \global\ftno=0
\def\foot{\global\advance\ftno by1\def\foot@rg{\hyperref{}{footnote}%
{\the\ftno}{\the\ftno}\xdef\foot@rg{\noexpand\hyperdef\noexpand\hypernoname%
{footnote}{\the\ftno}{\the\ftno}}}\footnote{$^{\foot@rg}$}}
%
%
%
\global\newcount\refno \global\refno=1
\newwrite\rfile
\def\ref{[\hyperref{}{reference}{\the\refno}{\the\refno}]\nref}
\def\nref#1{\DefWarn#1%
\xdef#1{[\noexpand\hyperref{}{reference}{\the\refno}{\the\refno}]}%
\writedef{#1\leftbracket#1}%
\ifnum\refno=1\immediate\openout\rfile=\jobname.refs\fi
\chardef\wfile=\rfile\immediate\write\rfile{\noexpand\item{[\noexpand\hyperdef%
\noexpand\hypernoname{reference}{\the\refno}{\the\refno}]\ }%
\reflabeL{#1\hskip.31in}\pctsign}\global\advance\refno by1\findarg}
\def\findarg#1#{\begingroup\obeylines\newlinechar=`\^^M\pass@rg}
{\obeylines\gdef\pass@rg#1{\writ@line\relax #1^^M\hbox{}^^M}%
\gdef\writ@line#1^^M{\expandafter\toks0\expandafter{\striprel@x #1}%
\edef\next{\the\toks0}\ifx\next\em@rk\let\next=\endgroup\else\ifx\next\empty%
\else\immediate\write\wfile{\the\toks0}\fi\let\next=\writ@line\fi\next\relax}}
\def\striprel@x#1{} \def\em@rk{\hbox{}}
\def\lref{\begingroup\obeylines\lr@f}
\def\lr@f#1#2{\DefWarn#1\gdef#1{\let#1=\UNd@FiNeD\ref#1{#2}}\endgroup\unskip}

\def\addref#1{\immediate\write\rfile{\noexpand\item{}#1}} 
\def\listrefs{\vfill\supereject\immediate\closeout\rfile\writestoppt
\baselineskip=\footskip\centerline{{\bf References}}\bigskip{\parindent=20pt%
\frenchspacing\escapechar=` \input \jobname.refs\vfill\eject}\nonfrenchspacing}
\def\startrefs#1{\immediate\openout\rfile=\jobname.refs\refno=#1}
\def\xref{\expandafter\xr@f}\def\xr@f[#1]{#1}
\def\refs#1{\count255=1[\r@fs #1{\hbox{}}]}
\def\r@fs#1{\ifx\UNd@FiNeD#1\message{reflabel \string#1 is undefined.}%
\nref#1{need to supply reference \string#1.}\fi%
\vphantom{\hphantom{#1}}{\let\hyperref=\relax\xdef\next{#1}}%
\ifx\next\em@rk\def\next{}%
\else\ifx\next#1\ifodd\count255\relax\xref#1\count255=0\fi%
\else#1\count255=1\fi\let\next=\r@fs\fi\next}
%

%
\newwrite\ffile\global\newcount\figno \global\figno=1
\def\fig{fig.~\hyperref{}{figure}{\the\figno}{\the\figno}\nfig}
\def\nfig#1{\DefWarn#1%
\xdef#1{fig.~\noexpand\hyperref{}{figure}{\the\figno}{\the\figno}}%
\writedef{#1\leftbracket fig.\noexpand~\xfig#1}%
\ifnum\figno=1\immediate\openout\ffile=\jobname.figs\fi\chardef\wfile=\ffile%
{\let\hyperref=\relax
\immediate\write\ffile{\noexpand\medskip\noexpand\item{Fig.\ %
\noexpand\hyperdef\noexpand\hypernoname{figure}{\the\figno}{\the\figno}. }
\reflabeL{#1\hskip.55in}\pctsign}}\global\advance\figno by1\findarg}
\def\xfig{\expandafter\xf@g}\def\xf@g fig.\penalty\@M\ {}
\def\figs#1{figs.~\f@gs #1{\hbox{}}}
\def\f@gs#1{{\let\hyperref=\relax\xdef\next{#1}}\ifx\next\em@rk\def\next{}\else
\ifx\next#1\xfig #1\else#1\fi\let\next=\f@gs\fi\next}
%
\def\figin{\epsfcheck\figin}\def\figins{\epsfcheck\figins}
\def\epsfcheck{\ifx\epsfbox\UnDeFiNeD
\message{(NO epsf.tex, FIGURES WILL BE IGNORED)}
\gdef\figin##1{\vskip2in}\gdef\figins##1{\hskip.5in}
\else\message{(FIGURES WILL BE INCLUDED)}%
\gdef\figin##1{##1}\gdef\figins##1{##1}\fi}
\def\DefWarn#1{}
\def\figinsert{\goodbreak\topinsert}
\def\ifig#1#2#3{\DefWarn#1\xdef#1{fig.~\the\figno}
\writedef{#1\leftbracket fig.\noexpand~\the\figno}%
\figinsert\figin{\centerline{#3}}
\smallskip
\leftskip=20pt \rightskip=20pt
\baselineskip12pt\noindent
{{\bf Fig.~\the\figno}\ \ninepoint #2}
\medskip
\global\advance\figno by1\par\endinsert}
\newwrite\lfile
{\escapechar-1\xdef\pctsign{\string\%}\xdef\leftbracket{\string\{}
\xdef\rightbracket{\string\}}\xdef\numbersign{\string\#}}
\def\writedefs{\immediate\openout\lfile=label.defs \def\writedef##1{%
{\let\hyperref=\relax\let\hyperdef=\relax\let\hypernoname=\relax
 \immediate\write\lfile{\string\def\string##1\rightbracket}}}}%
\def\writestop{\def\writestoppt{\immediate\write\lfile{\string\pageno
 \the\pageno\string\startrefs\leftbracket\the\refno\rightbracket
 \string\def\string\secsym\leftbracket\secsym\rightbracket
 \string\secno\the\secno\string\meqno\the\meqno}\immediate\closeout\lfile}}
\def\writestoppt{}\def\writedef#1{}

\def\seclab#1{\DefWarn#1%
\xdef #1{\noexpand\hyperref{}{section}{\the\secno}{\the\secno}}%
\writedef{#1\leftbracket#1}\wrlabeL{#1=#1}}
\def\subseclab#1{\DefWarn#1%
\xdef #1{\noexpand\hyperref{}{subsection}{\the\secno.\the\subsecno}%
{\the\secno.\the\subsecno}}\writedef{#1\leftbracket#1}\wrlabeL{#1=#1}}
\def\applab#1{\DefWarn#1%
\xdef #1{\noexpand\hyperref{}{appendix}{\secn@m}{\secn@m}}%
\writedef{#1\leftbracket#1}\wrlabeL{#1=#1}}
\newwrite\tfile \def\writetoca#1{}
\def\leaderfill{\leaders\hbox to 1em{\hss.\hss}\hfill}
\def\writetoc{\immediate\openout\tfile=\jobname.toc
   \def\writetoca##1{{\edef\next{\write\tfile{\noindent ##1
   \string\leaderfill{
   \string\hyperref{}{page}{\noexpand\number\pageno}%
   {\noexpand\number\pageno}} \par}}\next}}
}
\newread\ch@ckfile
\def\listtoc{\immediate\closeout\tfile\immediate\openin\ch@ckfile=\jobname.toc
\ifeof\ch@ckfile\message{no file \jobname.toc, no table of contents this pass}%
\else\closein\ch@ckfile\centerline{\bf Contents}\nobreak\medskip%
{\baselineskip=16pt\footnotefont\parskip=0pt\catcode`\@=11\input\jobname.toc
\catcode`\@=12\bigbreak\bigskip}\fi}
\catcode`\@=12 
\def\tenpoint{\def\rm{\fam0\tenrm}
\textfont0=\tenrm \scriptfont0=\sevenrm \scriptscriptfont0=\fiverm
\textfont1=\teni  \scriptfont1=\seveni  \scriptscriptfont1=\fivei
\textfont2=\tensy \scriptfont2=\sevensy \scriptscriptfont2=\fivesy
\textfont\itfam=\tenit \def\it{\fam\itfam\tenit}\def\footnotefont{\ninepoint}%
\textfont\bffam=\tenbf \def\bf{\fam\bffam\tenbf}\def\sl{\fam\slfam\tensl}\rm}
\font\ninerm=cmr9 \font\sixrm=cmr6 \font\ninei=cmmi9 \font\sixi=cmmi6
\font\ninesy=cmsy9 \font\sixsy=cmsy6 \font\ninebf=cmbx9
\font\nineit=cmti9 \font\ninesl=cmsl9 \skewchar\ninei='177
\skewchar\sixi='177 \skewchar\ninesy='60 \skewchar\sixsy='60
\def\ninepoint{\def\rm{\fam0\ninerm}
\textfont0=\ninerm \scriptfont0=\sixrm \scriptscriptfont0=\fiverm
\textfont1=\ninei \scriptfont1=\sixi \scriptscriptfont1=\fivei
\textfont2=\ninesy \scriptfont2=\sixsy \scriptscriptfont2=\fivesy
\textfont\itfam=\ninei \def\it{\fam\itfam\nineit}\def\sl{\fam\slfam\ninesl}%
\textfont\bffam=\ninebf \def\bf{\fam\bffam\ninebf}\rm}
%
\hyphenation{anom-aly anom-alies coun-ter-term coun-ter-terms}

\global\newcount\subsubsecno \global\subsubsecno=0
\def\subsubsec#1\par{\global\advance\subsubsecno by1%
{\toks0{#1}\message{(\the\secno\the\subsecno\the\subsubsecno. \the\toks0)}}%
\ifnum\lastpenalty>9000\else\bigbreak\fi
\noindent{\it\hyperdef\hypernoname{subsubsection}{\the\secno.\the\subsecno\the\subsubsecno}%
{\the\secno.\the\subsecno.\the\subsubsecno.} #1}
\par\nobreak\medskip\nobreak\noindent\ignorespaces}

\def\DefWarn#1{}
\def\tikzcaption#1#2{\DefWarn#1\xdef#1{Fig.~\the\figno}
\writedef{#1\leftbracket Fig.\noexpand~\the\figno}%
{
\smallskip
\leftskip=20pt \rightskip=20pt \baselineskip12pt\noindent
{{\bf Fig.~\the\figno}\ \ninepoint #2}
\bigskip
\global\advance\figno by1 \par}}

\def\ntoalpha#1{%
\ifcase#1%
@%
\or A\or B\or C\or D\or E\or F\or G\or H\or I
\fi
}

\global\newcount\appno \global\appno=1
\def\applab#1{\xdef #1{\ntoalpha\appno}\writedef{#1\leftbracket#1}\wrlabeL{#1=#1}
\global\advance\appno by1}

\def\preprint#1 #2\par{\rightline{\vbox{\baselineskip12pt\hbox{#1}\hbox{#2}}}\vskip2cm}
%
\def\title#1\par{\centerline{\bf #1}\nopagenumbers\pageno=0}
\def\author#1\par{\bigskip\bigskip\centerline{#1}}

\newcount\addressno

\def\email#1#2{\unskip$^#1$\footnote{\null}{\kern-\parindent \llap{$^#1$\hskip1pt}email: #2}}

\def\startcenter{%
  \par
  \begingroup
  \leftskip=0pt plus 1fil
  \rightskip=\leftskip
  \parindent=0pt
  \parfillskip=0pt
}
\def\stopcenter{\endgroup}

\def\address{\bigskip%
  \ifnum\the\addressno=0\else\stopcenter\endgroup\fi
  \advance\addressno by 1%
  \begingroup
  \startcenter
  \it
  \obeylines
  \addressAux
}
\def\addressAux#1{#1}

\def\abstract{\stopcenter\endgroup\bigskip\bigskip\noindent}

\def\Dsl{\,\raise.15ex\hbox{/}\mkern-13.5mu D} 
\def\dsl{\raise.15ex\hbox{/}\kern-.57em\partial}
 
\def\boxeqn#1{\vcenter{\vbox{\hrule\hbox{\vrule\kern3pt\vbox{\kern3pt
	\hbox{${\displaystyle #1}$}\kern3pt}\kern3pt\vrule}\hrule}}}

\def\lform{\hbox{$\sqcup$}\llap{\hbox{$\sqcap$}}}

\def\half{{1\over 2}}

\def\bar{\overline}
\def\({\left(}
\def\){\right)}



\def\qed{\hbox{\hskip 3pt
\vbox{\hrule\hbox to 7pt{\vrule height 7pt\hfill\vrule}
\hrule}}\hskip3pt}

\overfullrule=0pt\relax

\frenchspacing

\newread\instream \openin\instream= label.defs
\ifeof\instream \message{No labels in advance yet. Wait till next pass.}
\else \closein\instream \input label.defs
\fi
\writedefs

\def\arXiv:#1].{\hepthStrip#1 \nil}
\def\hepthStrip#1 #2\nil{\href{http://arxiv.org/abs/#1}{arXiv:#1 #2\unskip}].}

\input epsf
\input figflow
\input epsf

\def\centretable#1{ \hbox to \hsize {\hfill\vbox{
                    \offinterlineskip \tabskip=0pt \halign{#1} }\hfill} }

\preprint UUITP-13/23

\title Superspace Expansion of the 11D Linearized Superfields in 

\title The Pure Spinor Formalism, and The Covariant Vertex Operator

\author Maor Ben-Shahar\email{\dagger}{benshahar.maor@physics.uu.se} and Max Guillen\email{\ddagger}{max.guillen@physics.uu.se}

\address
Department of Physics and Astronomy, 75108 Uppsala, Sweden

\abstract
11D pure spinors have been shown to successfully describe 11D supergravity in a manifestly super-Poincar\'e covariant manner. The feasibility of its actual usage for scattering amplitude computations requires an efficient manipulation of the superfields defining linearized 11D supergravity. In this paper, we directly address this problem by finding the superspace expansions of these superfields, at all orders in $\theta$, from recursive relations their equations of motion obey in Harnad-Shnider-like gauges. After introducing the 11D analogue of the 10D ${\cal ABC}$ superparticle, we construct, for the first time, a fully covariant vertex operator for 11D supergravity by making use of the linearized 11D superfields. Notably, we show that this vertex reproduces the Green-Gutperle-Kwon 11D operators in light-cone gauge.
\Date {March 2023}

\newif\iffig
\figfalse



\lref\psstring{
N.~Berkovits,
``Super Poincare covariant quantization of the superstring,''
JHEP {\bf 04}, 018 (2000).
[arXiv:0001035 [hep-th]].
}

\lref\mafraone{
C.~R.~Mafra, O.~Schlotterer and S.~Stieberger,
``Complete N-Point Superstring Disk Amplitude I. Pure Spinor Computation,''
Nucl. Phys. B {\bf 873}, 419-460 (2013).
[arXiv:1106.2645 [hep-th]].
}

\lref\mafratwo{
C.~R.~Mafra, O.~Schlotterer and S.~Stieberger,
``Complete N-Point Superstring Disk Amplitude II. Amplitude and Hypergeometric Function Structure,''
Nucl. Phys. B {\bf 873}, 461-513 (2013).
[arXiv:1106.2646 [hep-th]].
}

\lref\mafrathree{
C.~R.~Mafra and O.~Schlotterer,
``Towards the n-point one-loop superstring amplitude. Part I. Pure spinors and superfield kinematics,''
JHEP {\bf 08}, 090 (2019).
[arXiv:1812.10969 [hep-th]].
}

\lref\mafrafour{
C.~R.~Mafra and O.~Schlotterer,
``Towards the n-point one-loop superstring amplitude. Part II. Worldsheet functions and their duality to kinematics,''
JHEP {\bf 08}, 091 (2019).
[arXiv:1812.10970 [hep-th]].
}

\lref\mafrafive{
C.~R.~Mafra and O.~Schlotterer,
``Towards the n-point one-loop superstring amplitude. Part III. One-loop correlators and their double-copy structure,''
JHEP {\bf 08}, 092 (2019).
[arXiv:1812.10971 [hep-th]].
}

\lref\mafrasix{
E.~D'Hoker, C.~R.~Mafra, B.~Pioline and O.~Schlotterer,
``Two-loop superstring five-point amplitudes. Part I. Construction via chiral splitting and pure spinors,''
JHEP {\bf 08}, 135 (2020).
[arXiv:2006.05270 [hep-th]].
}

\lref\mafraseven{
E.~D'Hoker, C.~R.~Mafra, B.~Pioline and O.~Schlotterer,
``Two-loop superstring five-point amplitudes. Part II. Low energy expansion and S-duality,''
JHEP {\bf 02}, 139 (2021).
[arXiv:2008.08687 [hep-th]].
}

\lref\pssuperparticle{
N.~Berkovits,
``Covariant quantization of the superparticle using pure spinors,''
JHEP {\bf 09}, 016 (2001).
[arXiv:0105050 [hep-th]].
}

\lref\bjornssonone{
J.~Bjornsson,
``Multi-loop amplitudes in maximally supersymmetric pure spinor field theory,''
JHEP {\bf 01}, 002 (2011).
[arXiv:1009.5906 [hep-th]].
}

\lref\bjornssontwo{
J.~Bjornsson and M.~B.~Green,
``5 loops in 24/5 dimensions,''
JHEP {\bf 08}, 132 (2010).
[arXiv:1004.2692 [hep-th]].
}

\lref\pssupermembrane{
N.~Berkovits,
``Towards covariant quantization of the supermembrane,''
JHEP {\bf 09}, 051 (2002).
[arXiv:0201151 [hep-th]].
}

\lref\maxparticle{
M.~Guillen,
``Equivalence of the 11D pure spinor and Brink-Schwarz-like superparticle cohomologies,''
Phys. Rev. D {\bf 97}, no.6, 066002 (2018).
[arXiv:1705.06316 [hep-th]].
}

\lref\maxone{
M.~Guillen,
``Pure spinors and $D=11$ supergravity,''
[arXiv:2006.06014 [hep-th]].
}

\lref\maxtwo{
M.~Guillen,
``Notes on the 11D pure spinor wordline vertex operators,''
JHEP {\bf 08}, 122 (2020).
[arXiv:2006.06022 [hep-th]].
}

\lref\maxthree{
M.~Guillen,
``Taming the 11D pure spinor b-ghost,''
[arXiv:2212.13653 [hep-th]].
}

\lref\maxfour{
N.~Berkovits, E.~Casali, M.~Guillen and L.~Mason,
``Notes on the $D=11$ pure spinor superparticle,''
JHEP {\bf 08}, 178 (2019).
[arXiv:1905.03737 [hep-th]].
}

\lref\maxberkovits{
N.~Berkovits and M.~Guillen,
``Equations of motion from Cederwall's pure spinor superspace actions,''
JHEP {\bf 08}, 033 (2018).
[arXiv:1804.06979 [hep-th]].
}

\lref\hsgauge{
J.~P.~Harnad and S.~Shnider,
``CONSTRAINTS AND FIELD EQUATIONS FOR TEN-DIMENSIONAL SUPERYANG-MILLS THEORY,''
Commun. Math. Phys. {\bf 106}, 183 (1986).
}

\lref\tsimpis{
D.~Tsimpis,
``Curved 11D supergeometry,''
JHEP {\bf 11}, 087 (2004).
[arXiv:0407244 [hep-th]].
}

\lref\greengutperlekwon{
M.~B.~Green, M.~Gutperle and H.~H.~Kwon,
``Light cone quantum mechanics of the eleven-dimensional superparticle,''
JHEP {\bf 08}, 012 (1999).
[arXiv:9907155 [hep-th]].
}

\lref\siegelabcparticle{
W.~Siegel,
``Classical Superstring Mechanics,''
Nucl. Phys. B {\bf 263}, 93-104 (1986).
}

\lref\brinkschwarz{
L.~Brink and J.~H.~Schwarz,
``Quantum Superspace,''
Phys. Lett. B {\bf 100}, 310-312 (1981).
}

\lref\equivalenceoperators{
A.~R.~Mikovic, C.~R.~Preitschopf and A.~E.~van de Ven,
``COVARIANT VERTEX OPERATORS FOR THE SIEGEL SUPERSTRING,''
Nucl. Phys. B {\bf 321}, 121-138 (1989).
}

\lref\equivalencestrings{
A.~R.~Mikovic and W.~Siegel,
``On-shell Equivalence of Superstrings,''
Phys. Lett. B {\bf 209}, 47-52 (1988).
}

\lref\failurequantizationabc{
A.~R.~Mikovic, M.~Rocek, W.~Siegel, P.~van Nieuwenhuizen, J.~Yamron and A.~E.~van de Ven,
``COVARIANTLY QUANTIZABLE SUPERPARTICLES,''
Phys. Lett. B {\bf 235}, 106-112 (1990).
}

\lref\covariantquantizationfirstilk{
F.~Essler, M.~Hatsuda, E.~Laenen, W.~Siegel, J.~P.~Yamron, T.~Kimura and A.~R.~Mikovic,
``Covariant quantization of the first ilk superparticle,''
Nucl. Phys. B {\bf 364}, 67-84 (1991).
}

\lref\brinkhowe{
L.~Brink and P.~S.~Howe,
``Eleven-Dimensional Supergravity on the Mass-Shell in Superspace,''
Phys. Lett. B {\bf 91}, 384-386 (1980).
}

\lref\maxdiegoone{
D.~G.~Sep\'ulveda and M.~Guillen,
``A Pure Spinor Twistor Description of Ambitwistor Strings,''
[arXiv:2006.06025 [hep-th]].
}

\lref\maxdiegotwo{
D.~Garc\'\i{}a Sep\'ulveda and M.~Guillen,
``A pure spinor twistor description of the $D = 10$ superparticle,''
JHEP {\bf 08}, 130 (2020).
[arXiv:2006.06023 [hep-th]].
}

\lref\mafraolimultitheta{
S.~Lee, C.~R.~Mafra and O.~Schlotterer,
``Non-linear gauge transformations in $D=10$ SYM theory and the BCJ duality,''
JHEP {\bf 03}, 090 (2016).
[arXiv:1510.08843 [hep-th]].
}

\lref\cederwallone{
M.~Cederwall,
``Towards a manifestly supersymmetric action for 11-dimensional supergravity,''
JHEP {\bf 01}, 117 (2010).
[arXiv:0912.1814 [hep-th]].
}

\lref\maxbghost{
N.~Berkovits and M.~Guillen,
``Simplified $D = 11$ pure spinor $b$ ghost,''
JHEP {\bf 07}, 115 (2017).
[arXiv:1703.05116 [hep-th]].
}

\lref\siegelpsone{
K.~Lee and W.~Siegel,
``Conquest of the ghost pyramid of the superstring,''
JHEP {\bf 08}, 102 (2005).
[arXiv:0506198 [hep-th]].
}

\lref\siegelpstwo{
K.~Lee and W.~Siegel,
``Simpler superstring scattering,''
JHEP {\bf 06}, 046 (2006).
[arXiv:0603218 [hep-th]].
}

\lref\FORMpackage{
J.~A.~M.~Vermaseren,
``New features of FORM,''
[arXiv:0010025 [math-ph]].
}

\lref\mafraPSS{
C.~R.~Mafra,
``PSS: A FORM Program to Evaluate Pure Spinor Superspace Expressions,''
[arXiv:1007.4999 [hep-th]].
}

\lref\mafrathesis{
C.~R.~Mafra,
``Superstring Scattering Amplitudes with the Pure Spinor Formalism,''
[arXiv:0902.1552 [hep-th]].
}

\lref\tsimpistend{
G.~Policastro and D.~Tsimpis,
``R**4, purified,''
Class. Quant. Grav. {\bf 23}, 4753-4780 (2006).
[arXiv:0603165 [hep-th]].
}

\lref\nathantwoloopone{
N.~Berkovits,
``Super-Poincare covariant two-loop superstring amplitudes,''
JHEP {\bf 01}, 005 (2006).
[arXiv:0503197 [hep-th]].
}

\lref\nathantwolooptwo{
N.~Berkovits and C.~R.~Mafra,
``Equivalence of two-loop superstring amplitudes in the pure spinor and RNS formalisms,''
Phys. Rev. Lett. {\bf 96}, 011602 (2006).
[arXiv:0509234 [hep-th]].
}

\lref\mafragomeztwoloop{
H.~Gomez and C.~R.~Mafra,
``The Overall Coefficient of the Two-loop Superstring Amplitude Using Pure Spinors,''
JHEP {\bf 05}, 017 (2010).
[arXiv:1003.0678 [hep-th]].
}

\lref\mafragomezthreeloop{
H.~Gomez and C.~R.~Mafra,
``The closed-string 3-loop amplitude and S-duality,''
JHEP {\bf 10}, 217 (2013).
[arXiv:1308.6567 [hep-th]].
}

\font\mbb=msbm10 
\newfam\bbb
\textfont\bbb=\mbb

\def\startcenter{%
  \par
  \begingroup
  \leftskip=0pt plus 1fil
  \rightskip=\leftskip
  \parindent=0pt
  \parfillskip=0pt
}
\def\stopcenter{%
  \par
  \endgroup
}

\listtoc
\writetoc
\filbreak

\newsec Introduction

\seclab\secone

 The pure spinor formalism for the superstring \psstring\ has been shown to be tremendously useful for efficient computation of scattering amplitudes involving bosonic and fermionic string states at tree- and loop-level \refs{\nathantwoloopone,\nathantwolooptwo,\mafragomeztwoloop,\mafragomezthreeloop}. Its respective field-theory limit, namely the 10D pure spinor superparticle \pssuperparticle, has also been proved to be convenient for studying and computing 10D super-Yang-Mills interactions, as well as for analyzing the high-energy behavior of the theory through the use of simple arguments based on zero mode counting and pure spinor algebraic properties \refs{\bjornssonone,\bjornssontwo}.

\medskip
\noindent Soon after the discovery of his new superstring formalism, Berkovits introduced the pure spinor versions of the 11D superparticle and supermembrane in \pssupermembrane. In this work, it was remarkably shown how the full field content of the Batalin-Vilkovisky description of linearized 11D supergravity can elegantly be described by the 11D pure spinor BRST cohomology. Although this fact gives the pure spinor formalism a privileged place as the appropriate framework for a consistent covariant quantization scheme in 11D, no explicit scattering amplitude computation has been carried out to date. This is mainly due to the lack of understanding of the building blocks needed for evaluating pure spinor correlators, including vertex operators of lower ghost numbers and 11D pure spinor identities.

\medskip
\noindent Over the past few years, this 11D pure spinor program has been revived, and some significant progress has been made. For instance, one of the authors recently introduced the ghost number one and two pure spinor vertex operators, and developed a new prescription for computing tree-level 11D pure spinor correlators \refs{\maxone,\maxtwo,\maxthree}. Likewise, some technical subtleties were found when trying to use a standard descent equation and define a ghost number zero vertex operator \maxfour, a fundamental piece for the calculation of four- and higher-point interactions in 11D supergravity. These results provide the toolbox needed for calculating three-particle scattering processes from pure spinor superspace expressions, and demand a revision or more careful analysis of the ghost number zero vertex operator.

\medskip
 In this paper we start the study of both issues mentioned above. As in 10D, the explicit computation of 11D pure spinor correlators requires the exact knowledge of the superspace expansions of all the superfields defining the 11D pure spinor vertex operators, namely the linearized 11D supergravity superfields. For this purpose, we find the complete set of equations of motion of linearized 11D supergravity in superspace, from the linearization of the 11D supergeometry and the four form field-strength of 11D supergravity. The use of Harnad-Shnider-like gauges \hsgauge\ on the lowest-dimensional components of the 11D superfields will be shown to give rise to a solvable system of recursive relations
 yielding every coefficient of the superspace expansions of all the linearized 11D superfields. The originality of our method relies on its feasibility and effectiveness within the pure spinor worldline framework. Indeed, our approach is pretty much exclusive and convenient for studying the specific forms of the 11D superfields involved in the construction of pure spinor vertex operators in 11D. This is in contrast to the general analysis carried out in \tsimpis, where superfields not directly relevant to the pure spinor formalism are studied. In this sense, the results of the first part of our paper will have a transcendent and direct significance for the development of the pure spinor program in 11D.

\medskip
\noindent The second part of our paper discusses the construction of a covariant vertex operator for 11D supergravity. This idea is strongly inspired by the relationship found between the vertex operators of the 10D ${\cal ABC}$ \siegelabcparticle\ and Brink-Schwarz \brinkschwarz\ superparticles in light-cone gauge \equivalenceoperators. The 11D light-cone gauge vertex operators were introduced by Green, Gutperle and Kwon in \greengutperlekwon. In order to reproduce these vertices from a covariant expression, we will first define the 11D analogue of the 10D ${\cal ABC}$ superparticle, and show it contains the same physical degrees of freedom as the standard 11D superparticle \greengutperlekwon. Next, we construct a covariant vertex operator 
by making exclusive use of supersymmetric quantities, as well as the linearized 11D superfields. The superspace expansions found in the first part of this work will allow us to show that this covariant operator exactly reproduces the Green-Gutperle-Kwon vertices in light-cone gauge.

\medskip
The paper is organized as follows. In section 2 we review the pure spinor formulation of the 11D superparticle, and discuss how the 11D supergravity physical states emerge from the cohomology of the pure spinor BRST charge. Section 3 motivates the definition of the linearized 11D superfields relevant to the definition of pure spinor vertex operators, and constructs the full set of equations of motion and gauge transformations satisfied by these. In section 4, we systematically solve the system of recursive relations found from the previous set of equations when superfields are subject to Harnad-Shnider-like gauges, and show they are self-consistent. Section 5 introduces the 11D ${\cal ABC}$ superparticle, and presents an 11D covariant vertex operator made out of supersymmetric worldline fields and the linearized 11D superfields, which is shown to reduce to the Green-Gutperle-Kwon vertex operators in light-cone gauge. Section 6 closes with discussions and future perspectives. We collect our conventions for gamma matrices in Appendix A, and briefly review the 10D ${\cal ABC}$ superparticle vertex operator and its relation to the light-cone gauge Brink-Schwarz operators in Appendix B. 

\newsec 11D Pure spinor superparticle 

\seclab\sectwo

\noindent The 11D pure spinor superparticle action is defined by \refs{\pssupermembrane,\maxparticle}
\eqnn \elevendpsaction
$$ \eqalignno{
S &= \int d\tau [P^a \partial_{\tau}X_a + p_{\alpha}\partial_{\tau}\theta^{\alpha} + w_{\alpha}\partial_{\tau}\lambda^{\alpha} - \half P^2] \ . & \elevendpsaction
}
$$
We use letters from the beginning of the Greek/Latin alphabet to denote spinor/vector $SO(1,10)$ indices. The variables $(P_a, p_{\alpha})$ are the conjugate momenta associated to the usual 11D superspace coordinates $(X^a, \theta^{\alpha})$. The bosonic spinor $\lambda^{\alpha}$ satisfies the 11D pure spinor constraint, i.e. $(\lambda\gamma^a \lambda) = 0$, and thus its respective conjugate momentum $w_{\alpha}$ is only defined up to the gauge transformation $\delta w_{\alpha} = (\gamma^{a}\lambda)_{\alpha}\sigma_a$, for any vector $\sigma_{a}$. Due to their wrong statistics, $(\lambda^{\alpha}, w_{\beta})$ will be referred to as ghost variables, and assigned to carry ghost charges 1 and -1, respectively. The 11D gamma matrices will be represented by $(\gamma^{a})_{\alpha\beta}$, $(\gamma^{a})^{\alpha\beta}$, and they satisfy the Clifford algebra: $(\gamma^{a})_{\alpha\beta}(\gamma^{b})^{\beta\delta} + (\gamma^{b})_{\alpha\beta}(\gamma^{a})^{\beta\delta} = 2\eta^{ab}\delta_{\alpha}^{\delta}$. We will raise and lower spinor indices by using the antisymmetric charge conjugation matrix $C_{\alpha\beta}$ and its inverse $C^{\alpha\beta}$, which obey the relation $C_{\alpha\beta}C^{\beta\delta} = \delta_{\alpha}^{\delta}$, so that $(\gamma^{a})^{\alpha\beta} = C^{\alpha\epsilon}C^{\beta\delta}(\gamma^{a})_{\epsilon\delta}$ for example (see Appendix A for more details).

\medskip
\noindent As is well-known, the space of physical states is defined by the cohomology of the BRST operator $Q = \lambda^{\alpha} d_{\alpha}$, where $d_{\alpha} = p_{\alpha} - \half (\gamma^a \theta)_{\alpha}P_{a}$ is the familiar primary constraint of the superparticle \greengutperlekwon. Such a cohomology can be shown to be non-trivial up to ghost number 7, describing the 11D supergravity states in its Batalin-Vilkovisky formulation. More explicitly, the ghost number 0, 1, 2 and 3 sectors respectively accommodate the gauge symmetry ghost-for-ghost-for-ghost; the gauge symmetry ghost-for-ghost; the supersymmetry, diffeomorphism and gauge symmetry ghosts; and the 11D supergravity physical fields. The higher ghost number sectors form a mirror of the fields described above, and correspond to the 11D supergravity antifields. One can easily see this by analyzing the ghost number three sector, $U^{(3)} = \lambda^{\alpha}\lambda^{\beta}\lambda^{\delta}A_{\alpha\beta\delta}$. The BRST-closedness condition implies that
\eqnn \closed
$$ \eqalignno{
Q\Psi &= 0  \rightarrow D_{(\alpha}A_{\beta\delta\epsilon)} = (\gamma^{a})_{(\alpha\beta}A_{a\delta\epsilon)} \ ,& \closed }
$$
and the BRST-exactness restriction imposes that
\eqnn \exact
$$ \eqalignno{
\delta \Psi &= Q\Lambda \rightarrow \delta A_{\alpha\beta\delta} = D_{(\alpha}\Lambda_{\beta\delta)} \ , & \exact
}
$$
where $\Lambda = \lambda^{\alpha}\lambda^{\beta}\Lambda_{\alpha\beta}$, and $\Lambda_{\alpha\beta}$ is a gauge parameter. These equations match the linearized equations of motion of 11D supergravity in superspace \brinkhowe, after making the identification $A_{\alpha\beta\delta} = C_{\alpha\beta\delta}$, where $C_{\alpha\beta\delta}$ is the linearized version of the lowest-dimensional component of the 11D supergravity super three form. As we will see later on, in a particular gauge, one can show that $U^{(3)}$ has the $\theta$-expansion,
\eqnn \psithetaexpansion
$$ \eqalignno{
U^{(3)} =&-{3\over 8}(\lambda \gamma^{b_1}\theta )(\lambda \gamma^{b_2}\theta )(\lambda \gamma^{b_1b_3}\theta )\epsilon^{b_2b_3}-{ 1 \over 8}(\lambda \gamma^{b_1}\theta )(\lambda \gamma^{b_2}\theta )(\lambda \gamma^{b_3}\theta )c^{b_1b_2b_3}\cr
&+{ 1 \over 5}(\lambda \gamma^{b_1}\theta )(\lambda \gamma^{b_2}\theta )(\lambda \gamma^{b_1b_3}\theta )(\theta \gamma^{b_3}\Psi^{b_2})-{ 1 \over 5}(\lambda \gamma^{b_1}\theta )(\lambda \gamma^{b_2}\theta )(\lambda \gamma^{b_3}\theta )(\theta \gamma^{b_1b_2}\Psi^{b_3})\cr
& + O(\theta^5) \ , &
\psithetaexpansion
}
$$
with $c_{abc}$, $\epsilon_{ab}$, $\Psi^a_{\alpha}$ being the three form, graviton and gravitino of 11D supergravity. Indeed, they can be shown to satisfy the linearized equations of motion 
\eqnn \equationsofmotion
$$\eqalignno{
\partial^{d}\partial_{[d}c_{abc]} &= 0 \ , \ \ \lform \epsilon_{bc} - 2 \partial^{a}\partial_{(b}\epsilon_{c)a} + \partial_{b}\partial_{c}(\eta^{ad}\epsilon_{ad}) = 0 \ , \ \ (\gamma^{abc})_{\alpha\beta}\partial_{b}\Psi_{c}^{\beta} = 0  \ ,& \equationsofmotion
}
$$
and gauge transformations
\eqnn \gaugetransformations
$$\eqalignno{
\delta c_{abc} &= \partial_{[a}s_{bc]} \ , \ \ \delta \epsilon_{ab} = \partial_{(a}t_{b)} \ , \ \ \delta \Psi_{a}^{\alpha} = \partial_{a}\kappa^{\beta} \ , & \gaugetransformations
}
$$
where $s_{ab}$, $t_{b}$ and $\kappa^{\beta}$ are arbitrary gauge parameters.

\medskip
\noindent As shown in \maxfour, it is also possible to describe the physical fields of linearized 11D supergravity through a ghost number one vertex operator involving momentum variables. Unlike the ghost number three  operator, this alternative operator describes the 11D supergravity three form gauge field through its field strength. Next, we review this construction and extend the analysis elaborated in \maxfour\ to find a complete set of superspace equations of motion giving rise to linearized 11D supergravity.

\newsec Linearized 11D supergravity equations of motion

\seclab\secthree

\noindent Let us first set some notation. We will use capital letters from the beginning/middle of the Latin alphabet to represent tangent/curved superspace indices, and lowercase letters from the beginning (middle) of the Latin/Greek alphabet to denote tangent (curved) space vector/spinor indices. The 11D supergeometry is then defined by the one form superfields $E^{A}$ and $\Omega_{B}{}^{C}$, referred to as the vielbein and spin-connection, and the super-Bianchi identities
\eqnn \elevendgeometry
$$
\eqalignno{
{\cal D}T^{A} = E^{B}R_{B}{}^{A} \ \ &, \ \ {\cal D}R_{A}{}^{B} = 0  \ , & \elevendgeometry\cr
}
$$
where $T^{A} = {\cal D}E^{A}$ is the super-torsion, $R_{A}{}^{B} = {\cal D}\Omega_{A}{}^{B}$ is the super-curvature, and ${\cal D} = E^{A}\nabla_{A}$ is the super-covariant derivative. Its action on an arbitrary tensor ${\cal F}_{A_{1}\ldots A_{m}}{}^{B_{1}\ldots B_{n}}$ is, 
\eqnn \covariantderivative
$$
\eqalignno{
{\cal D}{\cal F}_{A_{1}\ldots A_{m}}{}^{B_{1}\ldots B_{n}} = 
 d&{\cal F}_{A_{1}\ldots A_{m}}{}^{B_{1}\ldots B_{n}} + \Omega_{A_{1}}{}^{C}{\cal F}_{C A_{2}\ldots A_{m}}{}^{B_{1}\ldots B_{n}} + \ldots & \cr 
& - {\cal F}_{A_{1}\ldots A_{m}}{}^{C \ldots B_{n}}\Omega_{C}{}^{B_{1}} - \ldots \ ,&  \covariantderivative
}
$$
where $d$ is the ordinary exterior derivative. As is well-known, the following relations hold in Lorentz superspace
\eqnn \omegas
\eqnn \rs
$$
\eqalignno{
\Omega_{A \beta}{}^{\delta} &= {1\over 4}(\gamma^{bc})_{\alpha}{}^{\beta}\Omega_{A bc} \ ,& \omegas\cr 
R_{AB,\alpha}{}^{\beta} &= {1\over 4}(\gamma^{cd})_{\alpha}{}^{\beta}R_{AB,cd} \ .& \rs
}
$$

\subsec Review of the ghost number one vertex operator

\subseclab \secthreeone

\noindent As discussed in \maxfour, a simple way of defining a ghost number one vertex operator in the BRST-cohomology, is via a linear perturbation of the BRST charge $Q = \lambda^{\alpha}d_{\alpha}$, by $Q \rightarrow Q + U^{(1)}$. The nilpotency requirement of the deformed charge then automatically implies that $\{Q, U^{(1)}\} = 0$. This perturbation can readily be obtained from coupling the pure spinor superparticle \elevendpsaction\ to a curved background. When doing so, the BRST charge can be shown to be defined as $Q = \lambda^{\alpha}E_{\alpha}{}^{M}(P_{M} + \Omega_{M\beta}{}^{\delta}\lambda^{\beta}w_{\delta})$, where $P_{M}$ denotes the curved space supermomentum. Therefore, $U^{(1)}$ is given by
\eqnn \gnoneu
$$
\eqalignno{
U^{(1)} &= \lambda^{\alpha}(P_{a}h_{\alpha}{}^{a} + d_{\beta}h_{\alpha}{}^{\beta} - \Omega_{\alpha\beta}{}^{\delta}\lambda^{\beta}w_{\delta}) \ ,& \gnoneu
}
$$
where $h_{A}{}^{B} = \hat{E}_{A}{}^{M}E_{M}^{(1)B} = -E^{(1)M}_{A}\hat{E}_{M}{}^{B}$, ($\hat{E}_{A}{}^{M}$, $\hat{E}_{M}{}^{B}$) are the background values of the vielbeins, and ($E_{A}^{(1)M}$, $E_{M}^{(1)A}$) are their corresponding first order perturbations.

\medskip
\noindent As a check, one can explicitly compute $\{Q, U^{(1)}\} = 0$, to find the following relations
\eqnn \eomonea
\eqnn \eomtwoa
\eqnn \eomtwothree
$$ \eqalignno{
\lambda^{\alpha}\lambda^{\beta}P_{a}\bigg[D_{\alpha}h_{\beta}{}^{a} - h_{\alpha}{}^{\delta}(\gamma^{a})_{\beta\delta}\bigg] &= 0 \ , & \eomonea \cr
\lambda^{\alpha}\lambda^{\beta}d_{\delta}\bigg[D_{\alpha}h_{\beta}{}^{\delta} - \Omega_{\alpha\beta}{}^{\delta}\bigg] &= 0 \ ,& \eomtwoa \cr
\lambda^{\alpha}\lambda^{\beta}\lambda^{\delta}w_{\epsilon}R_{\alpha\beta,\delta}{}^{\epsilon} & = 0 \ . & \eomtwothree
}
$$
As we will see below, these equations become identities after plugging the superspace constraints of 11D supergravity.

\subsec Full set of equations of motion

\subseclab \secthreetwo

\noindent Eqns. \elevendgeometry\ imply the familiar relations
\eqnn \torsiondef
\eqnn \curvaturedef
$$
\eqalignno{
[\nabla_{A}, \nabla_{B}\} &= - T_{AB}{}^{C}\nabla_{C} - 2\Omega_{[AB\}}{}^{C}\nabla_{C} , & \torsiondef\cr
R_{AB,C}{}^{D} &= 2\nabla_{[A}\Omega_{B\}C}{}^{D} + T_{AB}{}^{F}\Omega_{FC}{}^{D} - \Omega_{[A|C|}{}^{F}\Omega_{B\}F}{}^{D}  \ , & \curvaturedef
}
$$
where $[ \, ,\,\}$ means graded commutator. The spectrum of 11D supergravity contains a three form gauge field which can be promoted to the three form superfield $F = E^{C}E^{B}E^{A}F_{ABC}$, satisfying the gauge transformation $\delta F = d L$, for any two form superfield $L$. Its field strength takes the form $G = dF$, and it satisfies the Bianchi identity $d G = 0$. In order to write down the full set of equations of motion of linearized 11D supergravity, one first expresses the covariant derivative $\nabla_{A} = E_{A}{}^{M}\partial_{M}$ at linear order as
\eqnn \introducingh
$$
\eqalignno{
\nabla_{A} &= D_{A} - h_{A}{}^{B}D_{B} \ ,& \introducingh 
}
$$
where $D_{A} = \hat{E}_{A}{}^{M}\partial_{M}$. The dynamical constraints $T_{\alpha\beta}{}^{a} = (\gamma^{a})_{\alpha\beta}$, $G_{\alpha\beta ab} = (\gamma_{ab})_{\alpha\beta}$, along with the conventional constraints $T_{\alpha\beta}{}^{\delta} = T_{a\alpha}{}^{c} = T_{ab}{}^{c} = G_{\alpha\beta\delta\epsilon} = G_{a\alpha\beta\delta} = G_{abc\alpha} = 0$ \brinkhowe, then imply the following set of equations of motion
\eqnn \eomone
\eqnn \eomtwo
\eqnn \eomthree
\eqnn \eomfour
\eqnn \eomfive
\eqnn \eomsix
$$
\eqalignno{
2D_{(\alpha}h_{\beta)}{}^{a} - 2h_{(\alpha}{}^{\delta}(\gamma^{a})_{\beta)\delta} + h_{b}{}^{a}(\gamma^{b})_{\alpha\beta} &= 0 \ ,& \eomone \cr
2D_{(\alpha}h_{\beta)}{}^{\delta} - 2\Omega_{(\alpha\beta)}{}^{\delta} + (\gamma^{a})_{\alpha\beta}h_{a}{}^{\delta} &= 0 \ ,& \eomtwo\cr
\partial_{a}h_{\alpha}{}^{\beta} - D_{\alpha}h_{a}{}^{\beta} - T_{a\alpha}{}^{\beta} - \Omega_{a\alpha}{}^{\beta} &= 0 \ ,& \eomthree\cr
\partial_{a}h_{\alpha}{}^{b} - D_{\alpha}h_{a}{}^{b} - h_{a}{}^{\beta}(\gamma^{b})_{\beta\alpha} + \Omega_{\alpha a}{}^{b} &= 0 \ , & \eomfour\cr
\partial_{a}h_{b}{}^{\alpha} - \partial_{b}h_{a}{}^{\alpha} - T_{ab}{}^{\alpha} &= 0 \ ,& \eomfive\cr
\partial_{a}h_{b}{}^{c} - \partial_{b}h_{a}{}^{c} - 2\Omega_{[ab]}{}^{c} &= 0 \ .& \eomsix 
}
$$
Notice that eqns. \eomone, \eomtwo\ immediately imply eqns. \eomonea, \eomtwoa, respectively. Using these constraints, one can also show that  (see \maxone\ for a detailed discussion) 
\eqnn \randtforelevendsugraone
\eqnn \randtforelevendsugratwo
\eqnn \randtforelevendsugrathree
$$
\eqalignno{
R_{(\alpha\beta,\delta)}{}^{\epsilon} + (\gamma^{a})_{(\alpha\beta}T_{a\delta)}{}^{
\epsilon} &= 0  \ ,& \randtforelevendsugraone\cr
R_{(\alpha\beta),b}{}^{c} + 2(\gamma^{c})_{\gamma(\beta}T_{|b|\alpha)}{}^{\gamma} &= 0 \ , & \randtforelevendsugratwo\cr
R_{(\alpha\beta),c}{}^{d} - 2D_{(\alpha}\Omega_{\beta)c}{}^{d} - (\gamma^{a})_{\alpha\beta}\Omega_{ac}{}^{d} &= 0  \ ,& \randtforelevendsugrathree
}
$$
where $T_{a\delta}{}^{\epsilon}$ is defined by the four form field strength $G$ via 
\eqnn \tadeltaepsilon
$$
\eqalignno{
T_{a\alpha}{}^{\beta} &= ({\cal T}_{a}{}^{bcde})_{\alpha}{}^{\beta}H_{bcde} \ , & \tadeltaepsilon
}
$$
and 
\eqnn \Tcal
$$
\eqalignno{
({\cal T}_{a}{}^{bcde})_{\alpha}{}^{\beta} &= {1\over 36}\bigg[\delta_{a}^{[b}(\gamma^{cde]})_{\alpha}{}^{\beta} + {1\over 8}(\gamma_{a}{}^{bcde})_{\alpha}{}^{\beta}\bigg] \ . & \Tcal
}
$$
Eqn. \randtforelevendsugraone\ immediately shows the validity of eqn. \eomtwothree.
\medskip
\noindent For later use, it will be convenient to rewrite $R_{\alpha\beta,b}{}^{c}$ in terms of $H_{abcd}$. This can readily be done through the use of eqn. \randtforelevendsugratwo. Explicitly,
\eqnn \randh
$$
\eqalignno{
R_{\alpha\beta,b}{}^{c} &= ({\cal R}_{b}{}^{cdefg})_{\alpha\beta}H_{defg} \ , & \randh
}
$$
with
\eqnn \Rcal
$$
\eqalignno{
({\cal R}_{bc}{}^{defg})_{\alpha\beta} &= {1\over 6}\bigg[\delta_{b}^{[d}\delta_{c}^{e}(\gamma^{fg]})_{\alpha\beta} + {1\over 24}(\gamma_{bc}{}^{defg})_{\alpha\beta}\bigg] \ . & \Rcal
}
$$
Likewise, eqn. \eomsix\ is automatically satisfied by the relation
\eqnn \omegaandh
$$
\eqalignno{
\Omega_{abc} &= \partial_{[a}h_{b]c} - \partial_{[a}h_{c]b} + \partial_{[c}h_{b]a} \ . & \omegaandh
}
$$

\medskip
\noindent
The equations of motion associated to the components of the linearized version of the three form superfield $F$ can directly be deduced from a four form superfield $H$ defined from the field strength $G$ as \maxberkovits
\eqnn \tensorcapitalh
$$
\eqalignno{
H_{ABCD} &= \hat{E}_{[D}{}^{Q}\hat{E}_{C}{}^{P}\hat{E}_{B}{}^{N}\hat{E}_{A\}}{}^{M}G_{MNPQ} \ , & \tensorcapitalh 
}
$$
which can equivalently be written as $H_{ABCD} = 4D_{[A}C_{BCD\}} + 6\hat{T}_{[AB}{}^{E}C_{ECD\}}$, where $C_{ABC} = \hat{E}_{[C}{}^{P}\hat{E}_{B}{}^{N}\hat{E}_{A\}}{}^{M}F_{MNP}$, and $\hat{T}^{A}$ is the flat space torsion. 
The expansion of \tensorcapitalh\ then yields
\eqnn \eomseven
\eqnn \eomeight
\eqnn \eomnine
\eqnn \eomten
$$
\eqalignno{
4D_{(\alpha}C_{\beta\delta\epsilon)} + 6(\gamma^{a})_{(\alpha\beta}C_{a\delta\epsilon)} &= 0 \ , & \eomseven\cr
 \partial_{a}C_{\alpha\beta\delta} - 3D_{(\alpha}C_{a\beta\delta)} + 3(\gamma^{b})_{(\alpha\beta}C_{ba\delta)}  &= -3(\gamma_{ab})_{(\alpha\beta}h_{\delta)}{}^{b} \ , & \eomeight \cr
2\partial_{[a}C_{b]\alpha\beta} + 2D_{(\alpha}C_{\beta) ab} + (\gamma^{c})_{\alpha\beta}C_{cab} &= 2(\gamma_{[b}{}^{c})_{\alpha\beta}h_{a]c} - 2(\gamma_{ab})_{\delta(\alpha}h_{\beta)}{}^{\delta} \ ,& \eomnine\cr
3\partial_{[a}C_{bc]\alpha} - D_{\alpha}C_{abc} & =  -3(\gamma_{[ab})_{\alpha\beta}h_{c]}{}^{\beta} \ . & \eomten
}
$$
The equations of motion displayed in \eomone-\eomsix\ are invariant under the gauge transformations
\eqnn \gtfull
$$
\eqalignno{
\delta h_{\alpha}{}^{a} = D_{\alpha}\Lambda^{a} + (\gamma^{a})_{\alpha\beta}\Lambda^{\beta} \ \ , \ \
\delta h_{\alpha}{}^{\beta} &= D_{\alpha}\Lambda^{\beta} + \Lambda_{\alpha}{}^{\beta} \ \ , \ \
\delta \Omega_{\alpha\beta}{}^{\epsilon} = D_{\alpha}\Lambda_{\beta}{}^{\epsilon} \ , &  \cr
\delta h_{a}{}^{b} = \partial_{a}\Lambda^{b} + \Lambda_{a}{}^{b} \ \ , \ \
\delta h_{a}{}^{\beta} &= \partial_{a}\Lambda^{\beta}\ \ , \ \ \delta \Omega_{a\alpha}{}^{\beta} = \partial_{a}\Lambda_{\alpha}{}^{\beta}  \ , & \gtfull
}
$$
where $\Lambda^{a}$, $\Lambda^{\alpha}$, $\Lambda_{\alpha}{}^{\beta} = {1\over 4}(\gamma^{ab})_{\alpha}{}^{\beta}\Lambda_{ab}$ are arbitrary gauge parameters. Similarly, the gauge transformations acting on the components of the superfield $C$, which leave the equations of motion listed in \eomseven-\eomten\ invariant, take the form
\eqnn \gtone
\eqnn \gttwo
\eqnn \gtthree
\eqnn \gtfour
$$
\eqalignno{
\delta C_{\alpha\beta\epsilon} &= D_{(\alpha}\Lambda_{\beta\epsilon)} + (\gamma^{a})_{(\alpha\beta}\Lambda_{a\epsilon)}  \ ,& \gtone\cr
\delta C_{a\alpha\epsilon} &= {1\over 3}\partial_{a}\Lambda_{\alpha\epsilon} + {2\over 3}D_{(\alpha}\Lambda_{\epsilon)a} + 
{1\over 3}(\gamma^{b})_{\alpha\epsilon}\Lambda_{ba} + (\gamma_{ab})_{\alpha\epsilon}\Lambda^{b}\ , & \gttwo\cr
\delta C_{ab\alpha} &= {2
\over 3}\partial_{[a}\Lambda_{b]\alpha} + {1\over 3}D_{\alpha}\Lambda_{ab} -(\gamma_{ab})_{\alpha\beta}\Lambda^{\beta} \ ,& \gtthree\cr
\delta C_{abc} &= \partial_{[a}\Lambda_{bc]}  \ .& \gtfour
}
$$
Next we use these transformations to conveniently fix the lowest-dimensional components of the $h$-, $\Omega$- and $C$-superfields to specific values. This gauge fixing will allow us to find a system of recursive relations, which will be systematically and explicitly solved to obtain the full $\theta$-expansions of the linearized 11D superfields.

\newsec Superspace expansions of the $h$-, $\Omega$- and $C$-superfields

\seclab\secthree

\noindent
In this section, we will show the system of equations defined by the relations \eomone-\eomsix\ and \eomseven-\eomten\ is closed, by explicitly solving it in the Harnad-Shnider-like gauges. 

\subsec Harnad-Shnider-like gauges

\subseclab\secthreeone

\noindent As done in 10D \hsgauge, one can use the gauge transformations listed in \gtfull-\gtfour\ to impose the so-called Harnad-Shnider-like gauges
\eqnn \hsgauge
$$
\eqalignno{
\theta^{\alpha}h_{\alpha}{}^{A} = 0 \ \ , \ \ \theta^{\alpha}\Omega_{\alpha A}{}^{B} &= 0  \ \ , \ \ \theta^{\alpha}C_{\alpha AB} = 0  \ .& \hsgauge
}
$$
After contracting both sides of the $C$-field equations of motion \eomseven-\eomten\ with $\theta^\alpha$, and introducing the $D$-operator $D = \theta^{\alpha}\partial_{\alpha}$, this gauge choice implies
\eqnn \eomsevenD
\eqnn \eomeightD
\eqnn \eomnineD
\eqnn \eomtenD
$$
\eqalignno{
(D+3)C_{\beta\delta\epsilon} + 3\theta^\alpha (\gamma^{a})_{\alpha(\beta}C_{a\delta\epsilon)} &= 0  \ ,& \eomsevenD\cr
   (D+2)C_{a\beta\delta} - 2\theta^\alpha(\gamma^{b})_{\alpha(\beta}C_{ba\delta)}  &= 2\theta^\alpha(\gamma_{ab})_{\alpha(\beta}h_{\delta)}{}^{b} \ ,& \eomeightD \cr
 (D+1)C_{\beta ab} + \theta^\alpha(\gamma^{c})_{\alpha\beta}C_{cab} &=  2\theta^\alpha(\gamma_{[b}{}^{c})_{\alpha\beta}h_{a]c} - \theta^\alpha(\gamma_{ab})_{\alpha\delta}h_{\beta}{}^{\delta} \ ,& \eomnineD\cr
 DC_{abc} & =  3(\theta \gamma_{[ab})_{\beta}h_{c]}{}^{\beta} \ .& \eomtenD
}
$$
Analogously, after contracting both sides of the equations of motion \eomone-\eomfour\ with $\theta^{\alpha}$, one gets for the $h$-fields
\eqnn \eomoneD
\eqnn \eomtwoD
\eqnn \eomthreeD
\eqnn \eomfourD
$$
\eqalignno{
(D+1)h_{\beta}{}^{a} - h_{\beta}{}^{\delta}(\theta\gamma^{a})_{\delta} + h_{b}{}^{a}(\theta\gamma^{b})_{\beta} &= 0  \ , & \eomoneD \cr
(D+1)h_{\beta}{}^{\delta} - {1\over 4}(\theta\gamma^{bc})^\delta\Omega_{\beta bc} + (\theta\gamma^{a})_{\beta}h_{a}{}^{\delta} &= 0  \ ,& \eomtwoD\cr
  Dh_{a}{}^{\beta} + 4 \theta^\alpha ({\cal T}_a{}^{bcde})_\alpha{}^\beta \partial_b C_{cde} + {1\over 4}(\theta\gamma^{bc})^\beta\Omega_{abc} &= 0 \ ,& \eomthreeD\cr
 Dh_{a}{}^{b} + (\theta\gamma^{b})_{\beta}h_{a}{}^{\beta} &= 0\ , & \eomfourD
}
$$
where we used eqns. \omegas, \tadeltaepsilon\ and  \omegaandh. Likewise, eqns. \randtforelevendsugratwo\ and \randtforelevendsugrathree\ together give
\eqnn \eomomegaD
$$
\eqalignno{
(1+D)\Omega_{\beta c}{}^d=(\theta{\cal R}_c{}^{d efgl})_\beta H_{efgl}-(\theta\gamma^a)_\beta\Omega_{ac}{}^d
&= 0 \ . & \eomomegaD
}
$$
The $\theta$-expansions of the superfields can now be obtained by recursively solving the equations above. The first step is to input the zeroth order in $\theta$ for the fields $C_{abc}$, $h_{a}{}^\alpha$ and $h_{ab}$,
\eqnn \eqnzeroththeta
$$
\eqalignno{
C_{abc} = c_{abc}+{\cal O}(\theta) \ , \ \ \ 
h_{a}{}^\alpha = -\Psi_a^\alpha +{\cal O}(\theta)\ , \ \ \ 
h_{ab} = -\epsilon_{ab} +{\cal O}(\theta) \ ,
&& \eqnzeroththeta
}
$$
then eqns. \eomtenD\ and \eomfourD\ can be used to find the ${\cal O}(\theta)$ terms in $C_{abc}$ and $h_{ab}$ respectively. Similarly, eqn. \eomfourD\ gives the $\theta^n$ term of $h_a{}^\alpha$ in terms of $\theta^{n-1}$ terms in $C_{abc}$ and $h_{ab}$. From these three fields the $\theta$-expansions of all other fields can be determined, as depicted in Figure 1. The interpretation of the figure is as follows: The $\theta^{n+1}$ terms in $\Omega_{\alpha ab}$ are determined by $\theta^{n}$ of $C_{abc}$ and $h_{ab}$. Then the newly determined $\Omega_{\alpha ab}$ and $h_a{}^\alpha$ give the components of $h_{\alpha}{}^{\beta}$. 
Similarly, the components of $h_\alpha{}^a$ and $C_{\alpha ab}$ are obtained from $h_{\alpha}{}^{\beta}$, $h_{ab}$, and $C_{abc}$.
Finally, $C_{\alpha ab}$ and $h_\alpha{}^a$ determine $C_{\alpha\beta a}$, which gives the superfield expansion of $C_{\alpha\beta\gamma}$. We will see this explicitly in the next section where we give the recursive equations and display the superfield expansions.

\ifig\figflowchart{Schematic representation of the equations of motion contracted with $\theta^\alpha$. Arrows indicate $\theta$-expansion dependency, for example, the arrow pointing from $C_{abc}$ to $\Omega_{\alpha ab}$ indicates that components of order $\theta^n$ in the former contribute to components of order $\theta^{n+1}$ in the latter.}
{\epsfxsize=.5\hsize\epsfbox{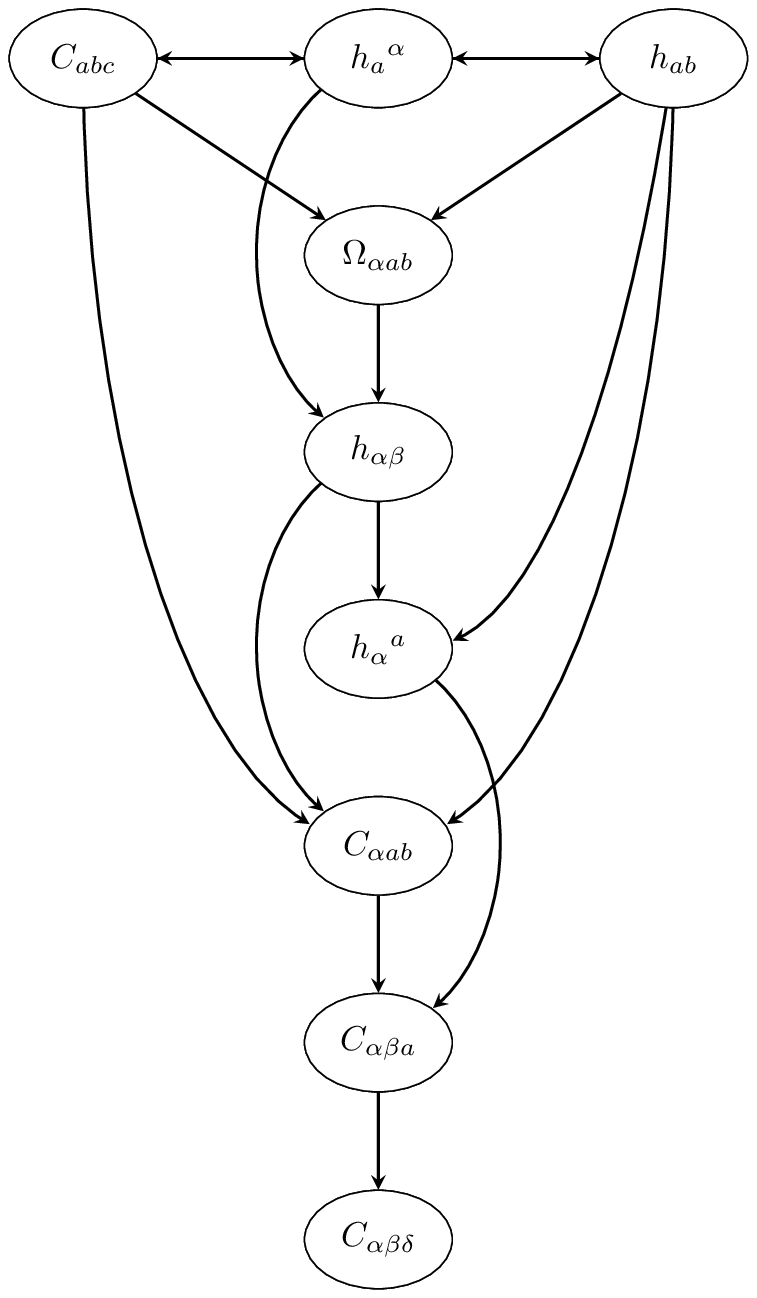}}

\subsec Example expansions

Here we provide the superfield expansion of $h_{a}{}^{b}$ and $h_{a}{}^{\alpha}$, which will be important in the next section where we provide a covariant vertex operator in 11D. In addition, we provide the ghost number three vertex operator for the pure spinor superparticle up to $\theta^5$. 
Starting with \eqnzeroththeta\ we can obtain the first order in $\theta$ of the higher-dimensional component of the $C$-field using eqn. \eomtenD. For higher orders in $\theta$, the $D$-operator just becomes a (non-zero) multiplicative factor and so it can be inverted. Then \eomthreeD, \eomfourD\ and \eomtenD\ can be solved, giving the initial set of recursion relations
\eqnn \recone
\eqnn \rectwo
\eqnn \recthree
$$
\eqalignno{
h_{a}{}^{\beta}\big|_{\theta^n} &= -{4\over n} \theta^\alpha ({\cal T}_a{}^{bcde})_\alpha{}^\beta \partial_b C_{cde}\big|_{\theta^{n-1}} -{1\over 4n}(\theta\gamma^{bc})^\beta\Omega_{abc}\big|_{\theta^{n-1}}\ ,& \recone\cr
 h_{a}{}^{b}\big|_{\theta^n} &= -{1\over n}(\theta\gamma^{b})_{\beta}h_{a}{}^{\beta}\big|_{\theta^{n-1}}\ , & \rectwo\cr
 C_{abc}\big|_{\theta^n} & =  {3\over n}\theta^\alpha( \gamma_{[ab})_{\alpha\beta}h_{c]}{}^{\beta}\big|_{\theta^{n-1}} \ .& \recthree
}
$$
The expansions of the remaining fields are then obtained from these initial three, using the remaining equations of motion. In order of dependence, they are
\eqnn \recfour
\eqnn \recfive
\eqnn \recsix
\eqnn \recseven
\eqnn \receight
\eqnn \recnine
$$
\eqalignno{
\Omega_{\alpha c}{}^{d}\big|_{\theta^{n}} &=   {4\over n+1} (\theta{\cal R}_c{}^{defgl})_\beta \partial_e C_{fgl}\big|_{\theta^{n-1}} -{1\over n+1} (\theta\gamma^a)_\alpha\Omega_{ac}{}^d\big|_{\theta^{n-1}}  \ ,& \recfour\cr
h_{\beta}{}^{\delta}\big|_{\theta^n} &= {1\over 4(n+1)} (\theta \gamma^{bc})^\delta \Omega_{\beta bc}\big|_{\theta^{n-1}} - {1\over n+1}(\theta\gamma^a)_\beta h_a{}^\delta\big|_{\theta^{n-1}}\ ,& \recfive\cr
h_{\beta}{}^{a}\big|_{\theta^n} &= {1\over n+1} (\theta \gamma^a)_\delta h_\beta{}^\delta\big|_{\theta^{n-1}}- {1\over n+1}(\theta\gamma^b)_\beta h_b{}^a\big|_{\theta^{n-1}}\ ,& \recsix\cr
C_{\beta ab}\big|_{\theta^n} &= {2\over n+1}(\theta \gamma_{[b}{}^c)_\beta h_{a]c}\big|_{\theta^{n-1}}-{1\over n+1}(\theta \gamma^c)_\beta C_{cab}\big|_{\theta^{n-1}}  - {1\over n+1} (\theta\gamma_{ab})_\delta h_\beta{}^\delta\big|_{\theta^{n-1}} \ \ \ \ \ & \recseven\cr
C_{a \beta \delta}\big|_{\theta^n} &= {2\over n+2} (\theta\gamma^b)_{(\beta}C_{\delta)ba}\big|_{\theta^{n-1}} + {2\over n+2} (\theta\gamma_{ab})_{(\beta}h_{\delta)}{}^b\big|_{\theta^{n-1}} \ ,& \receight\cr
C_{\beta\delta\epsilon}\big|_{\theta^n} &= -{3\over n+3} (\theta\gamma^a)_{(\beta}C_{\delta\epsilon)a}\big|_{\theta^{n-1}}  \ ,& \recnine
}
$$
We remind the reader that the tensors ${\cal T}$ and ${\cal R}$ can be found in eqns. \Tcal\ and \Rcal.
To simplify the expansions we replace $\partial_a\to k_a$, and introduce the Schoonschip notation, where vectors contracted with a tensor appear as indices. For example, $\gamma^a k_a\to \gamma^k$. Keeping terms up to $\theta^4$ in $h_{ab}$, we find the superfield expansion
\eqnn \habzero
\eqnn \habone
\eqnn \habtwo
\eqnn \habthree
\eqnn \habfour
$$
\eqalignno{
h^{a_1 a_2}\big|_{\theta^0} =
&-\epsilon^{a_1a_2} &\habzero\cr
h^{a_1 a_2}\big|_{\theta^1} =
&+(\theta \gamma^{a_2}\Psi^{a_1}) &\habone\cr
h^{a_1 a_2}\big|_{\theta^2} =
&+{ 1 \over 4}(\theta \gamma^{a_2b_1k}\theta )\epsilon^{a_1b_1}&\cr
&-{ 1 \over 2}(\theta {\cal T}^{a_1b_1b_2b_3b_4}\gamma^{a_2}\theta) h^{b_1b_2b_3b_4}&\habtwo\cr
h^{a_1 a_2}\big|_{\theta^3} =
&-{ 1 \over 24}(\theta \gamma^{a_2b_1k}\theta )(\theta \gamma^{a_1}\Psi^{b_1})&\cr
&-{ 1 \over 24}(\theta \gamma^{a_2b_1k}\theta )(\theta \gamma^{b_1}\Psi^{a_1})&\cr
&+{ 1 \over 24}(\theta \gamma^{a_2b_1b_2}\theta )(\theta \gamma^{b_1}\Psi^{b_2})k^{a_1}&\cr
&-{ 2 }(\theta {\cal T}^{a_1kb_1b_2b_3}\gamma^{a_2}\theta) (\theta \gamma^{b_1b_2}\Psi^{b_3}) &\habthree\cr
h^{a_1 a_2}\big|_{\theta^4} =
&-{ 1 \over 192}(\theta \gamma^{a_1b_1k}\theta )(\theta \gamma^{a_2b_2k}\theta )\epsilon^{b_1b_2}&\cr
&-{ 1 \over 192}(\theta \gamma^{a_2b_1k}\theta )(\theta \gamma^{b_1b_2k}\theta )\epsilon^{a_1b_2}&\cr
&+{ 1 \over 192}(\theta \gamma^{a_2b_1b_2}\theta )(\theta \gamma^{b_1b_3k}\theta )\epsilon^{b_2b_3}k^{a_1}&\cr
&-{ 1 \over 4}(\theta {\cal T}^{a_1kb_1b_2b_3}\gamma^{a_2}\theta) (\theta \gamma^{b_1b_2b_4k}\theta )\epsilon^{b_3b_4}&\cr
&-{ 1 \over 2}(\theta {\cal T}^{a_1kb_1b_2b_3}\gamma^{a_2}\theta) (\theta {\cal T}^{b_1b_4b_5b_6b_7}\gamma^{b_2b_3}\theta) h^{b_4b_5b_6b_7}&\cr
&+{ 1 \over 96}(\theta {\cal T}^{a_1b_1b_2b_3b_4}\gamma^{b_5}\theta) (\theta \gamma^{a_2b_5k}\theta )h^{b_1b_2b_3b_4}&\cr
&+{ 1 \over 96}(\theta {\cal T}^{b_1b_2b_3b_4b_5}\gamma^{a_1}\theta) (\theta \gamma^{a_2b_1k}\theta )h^{b_2b_3b_4b_5}&\cr
&+{ 1 \over 96}(\theta {\cal T}^{b_1b_2b_3b_4b_5}\gamma^{b_6}\theta) (\theta \gamma^{a_2b_1b_6}\theta )h^{b_2b_3b_4b_5}k^{a_1}&\habfour\cr
&+{\cal O}(\theta^5) \ ,
}
$$
and keeping terms up to $\theta^3$ for $h_a{}^\alpha$ we have
\eqnn \haalphazero
\eqnn \haalphaone
\eqnn \haalphatwo
\eqnn \haalphathree
$$
\eqalignno{
h_{a_1}^{\ \alpha}\big|_{\theta^0}=
&-\psi_a^\alpha &\haalphazero\cr
h_{a_1}^{\ \alpha}\big|_{\theta^1}=
&+{ 1 \over 2}(\gamma^{b_1k}\theta )^{\alpha}\epsilon^{a_1b_1}&\cr
&-(\theta {\cal T}^{a_1b_1b_2b_3b_4})^\alpha h^{b_1b_2b_3b_4}&\haalphaone\cr
h_{a_1}^{\ \alpha}\big|_{\theta^2}=
&-{ 1 \over 8}(\gamma^{b_1k}\theta )^{\alpha}(\theta \gamma^{a_1}\Psi^{b_1})&\cr
&-{ 1 \over 8}(\gamma^{b_1k}\theta )^{\alpha}(\theta \gamma^{b_1}\Psi^{a_1})&\cr
&+{ 1 \over 8}(\gamma^{b_1b_2}\theta )^{\alpha}(\theta \gamma^{b_1}\Psi^{b_2})k^{a_1}&\cr
&-6(\theta {\cal T}^{a_1kb_1b_2b_3})^\alpha(\theta \gamma^{b_1b_2}\Psi^{b_3}) &\haalphatwo\cr
h_{a_1}^{\ \alpha}\big|_{\theta^3}=
&-{ 1 \over 48}(\gamma^{b_1k}\theta )^{\alpha}(\theta \gamma^{a_1b_2k}\theta )\epsilon^{b_1b_2}&\cr
&-{ 1 \over 48}(\gamma^{b_1k}\theta )^{\alpha}(\theta \gamma^{b_1b_2k}\theta )\epsilon^{a_1b_2}&\cr
&+{ 1 \over 48}(\gamma^{b_1b_2}\theta )^{\alpha}(\theta \gamma^{b_1b_3k}\theta )\epsilon^{b_2b_3}k^{a_1}&\cr
&+{ 1 \over 24}(\gamma^{b_1k}\theta )^{\alpha}(\theta {\cal T}^{a_1b_2b_3b_4b_5}\gamma^{b_1}\theta) h^{b_2b_3b_4b_5}&\cr
&+{ 1 \over 24}(\gamma^{b_1k}\theta )^{\alpha}(\theta {\cal T}^{b_1b_2b_3b_4b_5}\gamma^{a_1}\theta) h^{b_2b_3b_4b_5}&\cr
&+{ 1 \over 24}(\gamma^{b_1b_2}\theta )^{\alpha}(\theta {\cal T}^{b_1b_3b_4b_5b_6}\gamma^{b_2}\theta) h^{b_3b_4b_5b_6}k^{a_1}&\cr
&-(\theta {\cal T}^{a_1kb_1b_2b_3})^\alpha(\theta \gamma^{b_1b_2b_4k}\theta )\epsilon^{b_3b_4}&\cr
&-{ 2 }(\theta {\cal T}^{a_1kb_1b_2b_3})^\alpha(\theta {\cal T}^{b_1b_4b_5b_6b_7}\gamma^{b_2b_3}\theta) h^{b_4b_5b_6b_7}&\haalphathree\cr
&+{\cal O}(\theta^4) \ .
}
$$
In addition we present the $\theta$-expansion of the ghost number three vertex operator $U^{(3)} = \lambda^\alpha\lambda^\beta\lambda^\gamma C_{\alpha\beta\gamma}$, which describes the physical fields of 11D supergravity in the cohomology of the pure spinor BRST charge \pssupermembrane. Up to $\theta^{5}$, it takes the form
\eqnn \clllthree
\eqnn \clllfour
\eqnn \clllfive
$$
\eqalignno{
C_{\lambda\lambda\lambda}\big|_{\theta^3}=
&-{3\over 8}(\lambda \gamma^{b_1}\theta )(\lambda \gamma^{b_2}\theta )(\lambda \gamma^{b_1b_3}\theta )\epsilon^{b_2b_3}&\cr
&-{ 1 \over 8}(\lambda \gamma^{b_1}\theta )(\lambda \gamma^{b_2}\theta )(\lambda \gamma^{b_3}\theta )c^{b_1b_2b_3}&\clllthree\cr
C_{\lambda\lambda\lambda}\big|_{\theta^4}=
&+{ 1 \over 5}(\lambda \gamma^{b_1}\theta )(\lambda \gamma^{b_2}\theta )(\lambda \gamma^{b_1b_3}\theta )(\theta \gamma^{b_3}\Psi^{b_2})&\cr
&-{ 1 \over 5}(\lambda \gamma^{b_1}\theta )(\lambda \gamma^{b_2}\theta )(\lambda \gamma^{b_3}\theta )(\theta \gamma^{b_1b_2}\Psi^{b_3}) &\clllfour\cr
C_{\lambda\lambda\lambda}\big|_{\theta^5}=
&+{ 1 \over 32}(\lambda \gamma^{b_1}\theta )(\lambda \gamma^{b_2}\theta )(\lambda \gamma^{b_1b_3}\theta )(\theta \gamma^{b_3b_4k}\theta )\epsilon^{b_2b_4}&\cr
&-{ 1 \over 32}(\lambda \gamma^{b_1}\theta )(\lambda \gamma^{b_2}\theta )(\lambda \gamma^{b_3}\theta )(\theta \gamma^{b_1b_2b_4k}\theta )\epsilon^{b_3b_4}&\cr
&-{11\over 192}(\lambda \gamma^{b_1}\theta )(\lambda \gamma^{b_2}\theta )(\lambda \gamma^{b_1b_3}\theta )(\theta {\cal T}^{b_2b_4b_5b_6b_7}\gamma^{b_3}\theta) h^{b_4b_5b_6b_7}&\cr
&-{11\over 192}(\lambda \gamma^{b_1}\theta )(\lambda \gamma^{b_2}\theta )(\lambda \gamma^{b_3}\theta )(\theta {\cal T}^{b_1b_4b_5b_6b_7}\gamma^{b_2b_3}\theta) h^{b_4b_5b_6b_7}&\cr
&-{ 1 \over 1536}(\lambda \gamma^{b_1}\theta )(\lambda \gamma^{b_2}\theta )(\lambda {\cal R}^{b_1b_2b_3b_4b_5b_6}\theta) (\theta \theta) h^{b_3b_4b_5b_6}&\cr
&+{ 1 \over 3072}(\lambda \gamma^{b_1}\theta )(\lambda \gamma^{b_1b_2}\theta )(\lambda {\cal R}^{b_3b_4b_5b_6b_7b_8}\theta) (\theta \gamma^{b_2b_3b_4}\theta )h^{b_5b_6b_7b_8}&\cr
&+{ 1 \over 3072}(\lambda \gamma^{b_1}\theta )(\lambda \gamma^{b_2}\theta )(\lambda {\cal R}^{b_3b_4b_5b_6b_7b_8}\theta) (\theta \gamma^{b_1b_2b_3b_4}\theta )h^{b_5b_6b_7b_8} &\clllfive\cr
&+{\cal O}(\theta^6) \ .
}
$$
The definitinos of the tensors ${\cal T}$ and ${\cal R}$ can be found in \Tcal\ and \Rcal. 
In these expansions we introduced the four-form field strength,
\eqnn \fieldstrength
$$
\eqalignno{
h_{abcd} &= 4\partial_{[a}c_{bcd]} \ , 
& \fieldstrength
}
$$
which makes gauge invariance under $c_{abc}\to \partial_{[a}\omega_{bc]}$ manifest. 
In an ancillary file to this paper we include expansions of all the fields up to $\theta^5$, both as they appear above and also after expanding the tensors ${\cal T}$, ${\cal R}$, and all resulting gamma matrix products\foot{Since the 3-point function in 11D pure spinor superspace involves the superfields $\Phi^{a} = \lambda^{\alpha}h_{\alpha}{}^{a}$ and $U^{(3)}$, see \refs{\maxone,\cederwallone,\maxberkovits}, we provide the $\theta$-expansion of $U^{(3)}$ up to $\theta^{7}$.}. These were obtained by implementing the recursion relations in FORM \FORMpackage\ and manipulating the gamma matrix products using routines presented in \mafraPSS.

\subseclab \secthreetwo

\noindent

\newsec A covariant vertex operator for 11D supergravity

\seclab \secfive
\noindent
This section introduces, for the first time, a covariant vertex operator for 11D supergravity in ordinary superspace. To this end, we first construct the 11D analogue of the ${\cal ABC}$ superparticle \siegelabcparticle.

\subsec The 11D ${\cal ABC}$ superparticle

\subseclab \secfiveone

\noindent 
It is well-known that the 11D superparticle possesses first- and second-class constraints, which cannot be easily separated out in a manifestly Lorentz covariant manner. As will be shown below (see Appendix B for the 10D analogue), one can overcome this difficulty by writing an alternative fully first-order framework, subject to a specific set of first-class constraints. The resulting theory, which we will refer to as the 11D ${\cal ABC}$ superparticle, will then be shown to be physically equivalent to the original 11D superparticle \greengutperlekwon.

\medskip
\noindent The 11D ${\cal ABC}$ superparticle action will be defined as
\eqnn \abcsuperparticle
$$
\eqalignno{
S &= \int d\tau [P^{a}\partial_{\tau}X_{a} + p_{\alpha}\partial_{\tau}\theta^{\alpha} + \rho {\cal A} + \xi_{\alpha}{\cal B}^{\alpha} + \iota^{\alpha\beta}{\cal C}_{\alpha\beta}]
& \abcsuperparticle
}
$$
where $\rho$, $\xi_{\alpha}$, $\iota^{\alpha\beta}$ are the Lagrange multipliers associated to the constraints 
\eqnn \elevendabcconstraints
$$
\eqalignno{
{\cal A} = P^{a}P_{a} \ \ , \ \ {\cal B}^{\alpha} &= (\gamma^{a}d)^{\alpha}P_{a} \ \ , \ \ {\cal C}_{\alpha\beta} = d_{[\alpha}d_{\beta]} & \elevendabcconstraints
}
$$
and $d_{\alpha}$ is defined as in section 2. The only non-zero (anti)commutators describing the constraint algebra are given by
\eqnn \constraintalgebrabac
$$
\eqalignno{
\{{\cal B}^{\alpha}, {\cal B}^{\beta}\} = -(\gamma^{a})^{\alpha\beta}P_{a}{\cal A} \ \ &, \ \ [{\cal C}_{\alpha\beta}, {\cal C}_{\delta\epsilon}] = -4(
\gamma^{a})_{\underline{\beta}\bar{\delta}}P_{a}{\cal C}_{\underline{\alpha}\bar{\epsilon}} \ \ , \ \ [{\cal B}^{\alpha},{\cal C}_{\beta\delta}] = 2\delta^{\alpha}_{[\delta}d_{\beta]}{\cal A} & \cr
& & \constraintalgebrabac
}
$$
where we are using barred and underlined letters to denote index antisymmetrization. 
 In order to show that the physical degrees of freedom of the model \abcsuperparticle\ match those of the 11D superparticle, we need to define the so-called light-cone gauge.

\subsec Light-cone gauge

\subseclab \secfivetwo

\noindent 
We begin by defining the light-cone directions 
\eqnn \lcdirections
$$
\eqalignno{
X^{\pm} &= {1\over \sqrt{2}}(X^{0}\pm X^{10})\ ,& \lcdirections
}
$$
and transverse directions are denoted by $i,j,l$. We take gamma matrices to be represented by
\eqnn \gammamatrices
$$
\eqalignno{
\gamma^{i}_{\alpha\beta}= 
\left(\matrix{
    0 & \sigma^i_{A \dot{A}} \cr
    -\sigma^i_{\dot{B}B} & 0 
}\right) ,
\ \ 
\gamma^{+}_{\alpha\beta} = 
\left(\matrix{
    0 & 0 \cr
    0 & -i\sqrt{2}I_{\dot{B}\dot{A}}
}\right) ,
\ \
\gamma^{-}_{\alpha\beta} = 
\left(\matrix{
    -i\sqrt{2}I_{AB} & 0 \cr
    0 & 0
}\right)
 ,& & \gammamatrices
}
$$
where $\sigma^{i}$ are $SO(9)$ Pauli matrices and $A, \dot{A}$ are $SO(9)$ spinor indices. Although we use dotted index notation as in 10D, in 11D these indices can be contracted using the charge conjugation matrix,
\eqnn \cmatrix
$$
\eqalignno{
C_{\alpha\beta} &= 
\left(\matrix{
    0 & -I_{\dot{A}A} \cr
    I_{B\dot{B}} & 0 
}\right)\ . & \cmatrix
}
$$
As usual, one can use the constraint ${\cal A}=P^aP_a$ to fix $X^{+}$, and to determine $P^{-}$,
\eqnn \aconstraintlc
$$
\eqalignno{
X^{+} = x^{+} + \tau P^{+} \ &,  \ \ P^{-} = {P^{i}P^{i}\over 2P^{+}}  \ .& \aconstraintlc
}
$$
Moreover, one can use the constraint ${\cal B}^{\alpha}$ to fix one of the $SO(9)$ components of $\theta^{\alpha}$ and $d_{\alpha}$
\eqnn \bconstraintlc
$$
\eqalignno{
(\gamma^{+}\theta)_{\alpha} = 0 \ \ &,  \ \ (\gamma^{-}d)^{\alpha} = {1\over P^{+}}[-(\gamma^{+}d)^{\alpha}P^{-} + (\gamma^{i}d)^{\alpha}P_{i}] \ .& \bconstraintlc
}
$$
The remaining variables are then given by $(X^{-}, X^{i}, P^{+}, P^{i}, \theta^{A}, p_{B})$. For convenience, instead of $(\theta^{A}, p_{B})$ we will use the pair of variables $(d^{A}, q_{B})$, where $q_{\alpha} = p_{\alpha} + {1\over 2}(\gamma^{a}\theta)_{\alpha}P_{a}$ is the supersymmetric charge. It is not hard to check that,
\eqnn \qdalgebras
$$
\eqalignno{
\{d_{\alpha}, d_{\beta}\} = -(\gamma^{a})_{\alpha\beta}P_{a} \ \ , \ \ \{d_{\alpha}, q_{\beta}\} &= 0 \ \ , \ \ \{q_{\alpha}, q_{\beta}\} = (\gamma^{a})_{\alpha\beta}P_{a} \ , & \qdalgebras
}
$$
which implies that
\eqnn \qdalgebraslc
$$
\eqalignno{
\{d_{A}, d_{B}\} = -\delta_{AB} \ \ , \ \ \{d_{A}, q_{B}\} &= 0 \ \ , \ \ \{q_{A}, q_{B}\} = \delta_{AB} \ , & \qdalgebraslc
}
$$
where we used the redefinition $d_{A} \rightarrow \sqrt{\sqrt{2}iP^{+}} d_{A}$. In this manner, the constraint ${\cal C}_{\alpha\beta}$ requires that $d_{A}$ obeys the following gauge transformation,
\eqnn \calphabetad
$$
\eqalignno{
\delta d_{A} &= m_{A}{}^{B}d_{B} \ ,& \calphabetad
}
$$
where $m_{AB}$ is a completely antisymmetric matrix. The constraint ${\cal C}_{\alpha\beta} = 0$ requires that $d_{A} = \chi y_{A}$, where $\chi$ is a fermionic constant, and $y_{A}$ is a bosonic $SO(9)$ spinor, and so eqn. \calphabetad\ allows us to set $d_{A} = (d_{1}, 0, \ldots, 0)$. Using that ${\cal C}_{\alpha\beta}$ is invariant under $d_{\alpha} \rightarrow -d_{\alpha}$, one can fix the eigenvalue of $d_{1}^{2} = -1$, and therefore the only dynamical variables are defined by $(X^{-}, X^{i}, P^{+}, P^{i}, q_{A})$. This is exactly the same number and type of variables describing the light-cone gauge 11D superparticle \greengutperlekwon. Indeed, the Hilbert space is spanned by the vector space realizing the $q$-algebra in \qdalgebraslc, i.e. it is described by $2^{{16\over 2}} = 256$ states, the number of the 11D supergravity physical states.

\subsec Vertex operator

\subseclab \secfivethree

\noindent  The covariant vertex operator will be made out of supersymmetric quantities, and the linearized 11D supergravity superfields studied in previous sections. As in 10D, see Appendix B for a short review, after adequately imposing the light-cone gauge conditions and solving the ${\cal C}$-constraint of \elevendabcconstraints, the covariant operator will be shown to coincide with the Green-Gutperle-Kwon vertices from the 11D superparticle \greengutperlekwon.

\medskip
\noindent Concretely, we define the 11D covariant vertex operator as
\eqnn \covop
$$
\eqalignno{
V &= P^{a}P^{b}h_{ab} + P^{a}h_{a}{}^{\alpha}d_{\alpha} \ ,& \covop
}
$$
where $h_{ab}$, $h_{a}{}^{\alpha}$ are the linearized superfields of section 3. This vertex is the 11D analogue of the Siegel vertex operator for 10D super-Yang-Mills \siegelabcparticle, see eqn. B.2.

\noindent In addition to the constraints fixed for the world-line variables in the previous section, we also need to gauge fix the physical fields. We take light-cone gauge fixing conditions
\eqnn \lchfield
$$
\eqalignno{
\epsilon_a{}^+ = 0 \ , \ \ \ \ 
\Psi^{\alpha+} &= 0 \ , \ \ \ \ 
c_{ab}{}^{+} = 0 \ .
& \lchfield
}
$$
We assume that the momentum carried by a physical state satisfies $k^+=0$, and that $k^-$ is non-infinite. This means that the $k^i$ components need to be complex, in order to maintain the massless condition $k^2=0$. Using residual gauge freedom and the conditions in \lchfield\ we can further fix
\eqnn \lcgaugefieldstwo
$$
\eqalignno{
\epsilon_{ij}k^i &= 0 \ , \ \ \ \ 
\epsilon_i{}^i = 0 \ . & \lcgaugefieldstwo
}
$$
For the three form $c_{abc}$ we have simply 
\eqnn \lccfield
$$
\eqalignno{
c_{iab}k^i  &= 0 \ .& \lccfield
}
$$
Additional constraints can be imposed on the gravitino but for this section we only focus on the bosonic sector, so we ignore it from now. 

\medskip
\noindent To proceed with the comparison to \greengutperlekwon\ we sectorize the vertex operator into parts containing $\epsilon^{--},\ \epsilon^{-i},\ \epsilon^{ij},\ c_{ij}{}^-$ and $c_{ijl}$. Keeping only terms with $\epsilon^{--}$ and inserting the superfield expansions of the previous section into the vertex operator \covop\ we find that all terms containing $\theta$ vanish due to the light-cone gauge conditions and the fact that the one-form $\gamma^a$ and two-form $\gamma^{ab}$ are symmetric in their spinor indices. For example, we have
\eqnn \threeform
$$
\eqalignno{
(\theta\gamma^{a+b}\theta)&= {1\over 3}(\theta\gamma^{a}\gamma^+\gamma^b\theta) = 0 \ , & \threeform
}
$$
and similarly 
\eqnn \fourform
$$
\eqalignno{
(\theta\gamma^{a+bc}\theta)& = 0 \ . & \fourform
}
$$
Due to this, the vertex operator for $\epsilon^{--}$ becomes
\eqnn \vohmm
$$
\eqalignno{
V\big|_{\epsilon^{--}}& = -P^+P^+ \epsilon^{--} \ . & \vohmm
}
$$
For the $\epsilon^{i-}$ components of the graviton, one also needs to make use of 
\eqnn \thetaid
$$
\eqalignno{
(\theta \gamma^{i_1\ldots i_n}\theta) & \propto (\theta \gamma^{i_1\ldots i_n}\gamma^+\gamma^-\theta)  = 0 \ , & \thetaid
}
$$
as well as the substitution
\eqnn \thetatoqd
$$
\eqalignno{
\theta^\alpha &= {1\over 2P^+}(\gamma^+)^{\alpha\beta}(q_\beta-d_\beta) \ , & \thetatoqd
}
$$
which are both valid in light-cone gauge. The vertex operator then becomes 
\eqnn \vohimone
$$
\eqalignno{
V\big|_{\epsilon^{i-}}& = \epsilon^{i-}P_iP^+-{1\over 16}(q\gamma^{+ik}q) \epsilon^{i-}+{1\over 16}(d\gamma^{+ik_1}d) \epsilon^{i-} \cr
&=\epsilon^{i-}P_iP^+-{1\over 16}(q\gamma^{+ik}q) \epsilon^{i-} \ . & \vohimone
}
$$
The second equality comes about because the term with two $d_\alpha$ vanishes due to the $C_{\alpha\beta}$ constraint. In fact, any term with more than one $d_\alpha$ will automatically vanish by using Fierz identities and the $C_{\alpha\beta}$ constraint in \elevendabcconstraints.
Next, the term containing two $q_\alpha$ charges can be mapped to $SO(9)$ rotation generators introduced in \greengutperlekwon. Defining
\eqnn \Rij
$$
\eqalignno{
R^{ij}& = {1\over 16P^+}(q\gamma^{+ij}q) \ , & \Rij
}
$$
they obey the algebra
\eqnn \Rijalgebra
$$
\eqalignno{
[R^{ij},R_{lk}]& = \delta^{[i}_{[k}\delta^{j]}_{l]} \ , & \Rijalgebra
}
$$
and in terms of these generators the final form of the vertex operator is,
\eqnn \vohimone
$$
\eqalignno{
V\big|_{\epsilon^{i-}} &=\epsilon^{i-}P_iP^+-R^{ik}P^+ \epsilon^{i-} \ . & \vohimone
}
$$
For the transverse components $\epsilon_{ij}$ we use the identities \threeform , \fourform, \thetaid , as well as the replacement for $\theta$ variables in terms of $q$ and $d$ with \thetatoqd\ to find the vertex operator
\eqnn \vohijone
$$
\eqalignno{
V\big|_{\epsilon^{ij}} &=
-\epsilon_{ij}P^iP^j+{1\over 8} (q\gamma^{+ik}q)\epsilon_{ij}P^j (P^+)^{-1}
\cr&
+{1\over 768} (q\gamma^{+aik}q)(q\gamma^{+ajk}q)\epsilon_{ij}(P^+)^{-2}
-{1 \over 768} (q \gamma^{+i k}q)(q\gamma^{+j k}q)\epsilon_{ij}(P^+)^{-2}
\ . 
& \vohijone
}
$$
For $SO(9)$ we have the Fierz identity
\eqnn \soninefierz
$$
\eqalignno{
q_A q_B &= q^2\delta_{AB} + {1\over 32}(\sigma_{ij})_{AB} (q\sigma^{ij}q) + {1\over 96}(\sigma_{ijl})_{AB}(q\sigma^{ijl}q)
\ ,
& \soninefierz
}
$$
It is not hard to show that this equation implies the identity
\eqnn \threetotwoforms
$$
\eqalignno{
(q\gamma^{+(i|ja}q)(q\gamma^{+|l)jb}q)k_{a} k_{b} h_{il} &= 5 (q\gamma^{+(i|a}q)(q\gamma^{+|l)b}q)k_{a} k_{b} h_{il} 
\ ,
& \threetotwoforms
}
$$
To show this, recall that $\gamma^+$ is a projector, and that the expressions above are proportional to $SO(9)$ spinor expressions, for example $(q\gamma^{+ijl}q)\propto (q\sigma^{ijl}q)$. Plugging eqn. \threetotwoforms\ in \vohijone\ then gives the result
\eqnn \vohijone
$$
\eqalignno{
V\big|_{\epsilon^{ij}} &=
-\epsilon_{ij}P^iP^j+{1\over 8} (q\gamma^{+ik}q)\epsilon_{ij}P^j (P^+)^{-1}
-{1\over 128}  (q \gamma^{+i k}q)(q\gamma^{+j k}q)\epsilon_{ij}(P^+)^{-2}
\ , \cr
&=
-\epsilon_{ij}P^iP^j+ 2 R^{ik}\epsilon_{ij}P^j
-2 R^{ik}R^{jk}\epsilon_{ij} \ ,
& \vohijone
}
$$
which is in complete agreement with \greengutperlekwon.
\medskip
\noindent
For the three form vertex operators we work along similar lines and find
\eqnn \vocijm
\eqnn \vocijl
$$
\eqalignno{
V\big|_{c_{ij}{}^-} &=
{1\over 96}(q\gamma^{+ijl}q)H_{ijl}{}^{-} \ ,& \vocijm \cr
V\big|_{c_{ijl}} &=
H_ {i_ 1 i_ 2 i_ 3 i_ 4}
\Big(P^{i_1} - {1\over 24}(q\gamma^{+i_1 k} q)\Big){1\over 96} (q\gamma^{+ i_2 i_3 i_4} q)
 \ .& \vocijl
}
$$
Identifying $R^{ijl} = {1\over 96}(q\gamma^{+ijl}q)$, these components of the vertex operator become
\eqnn \vocijmR
\eqnn \vocijlR
$$
\eqalignno{
V\big|_{c_{ij}{}^-} &=
R^{ijl}H_{ijl}{}^{-} \ ,& \vocijmR \cr
V\big|_{c_{ijl}} &=
H_ {i_ 1 i_ 2 i_ 3 i_ 4}
\Big(P^{i_1} - {2\over 3}R^{i_1 k}\Big)R^{i_2 i_3i_4}
 \ ,& \vocijlR
}
$$
which is once again in agreement with \greengutperlekwon.

The analysis for the fermionic states immediately follows from supersymmetry arguments.

\newsec Discussions

\seclab \secfive

\noindent In this work, we have found a compact and straightforward list of recursive relations, \recone-\recnine\ , which determine the superspace expansions of all the superfields describing linearized 11D supergravity, and which are relevant to the pure spinor formalism. These results possess a variety of applications including the computation of three-particle interactions with manifest supersymmetry, the construction of a new pure spinor twistor transform describing 11D supergravity along the lines of \refs{\maxdiegoone,\maxdiegotwo,\maxbghost}, the superspace expansion of multiparticle superfields relevant to perturbiner methods\foot{See \mafraolimultitheta\ for the 10D analogue of this statement.}, among others. We plan to explore these directions further in the near future. In particular, the 3-point correlator has been found in \cederwallone\ from a field-theory perspective, and in \maxone\ from a worldline approach. The component amplitudes should then immediately follow from an appropriate projection procedure, and the $\theta$-expansions presented in this paper. We plan to tackle this problem in the near future, as well as to extend the state-of-the-art amplitude prescription, so that supergravity interactions involving an arbitrary number of external bosonic and fermionic states could directly be obtained from pure spinor superspace expressions.

\medskip
Furthermore, we have introduced the so-called 11D ${\cal ABC}$ superparticle, and have shown that the respective covariant vertex operator \covop\ reduces to the Green-Gutperle-Kwon operators after imposing the light-cone gauge. This result also gives rise to several follow-up ideas. For instance, although the 10D ${\cal ABC}$ superparticle is equivalent to the Brink-Schwarz superparticle in light-cone gauge \equivalencestrings, its covariant quantization fails in describing 10D super-Yang-Mills \failurequantizationabc. One possible way of fixing this issue is by introducing an extra constraint, which defines the so-called first-ilk or ${\cal ABCD}$ superparticle. It is the BRST quantization of this model which reproduces the right physical spectrum \covariantquantizationfirstilk. It would be interesting to investigate if the same phenomenon occurs in 11D, as well as to explore the possible modifications one needs to make to the 11D ${\cal ABC}$ superparticle here proposed, so that the BRST treatment of the resulting model reproduces the physical degrees of freedom of 11D supergravity. It is worthwhile mentioning that the infinite tower of ghosts in the BRST-closed vertex operator of the 10D ${\cal ABCD}$ superparticle can effectively be described by the pure spinor sector of the ghost number zero vertex operator in the pure spinor worldline formalism \refs{\siegelpsone,\siegelpstwo}. This means that the computation of the BRST-closed operator in the 11D ${\cal ABCD}$ superparticle will provide extremely important information about the structure of the ghost sector in the pure spinor vertex operator of ghost number zero. We leave the study of this issue and related topics for future work.

\bigskip \noindent{\bf Acknowledgements:} We are grateful to Nathan Berkovits and Carlos Mafra for providing valuable comments on an early draft of this work. We would also like to thank Oliver Schlotterer and Henrik Johansson for useful discussions on related topics. This research was supported in part by the Knut and Alice Wallenberg Foundation under grants KAW 2018.0116 ({\it From Scattering Amplitudes to Gravitational Waves}) and KAW 2018.0162 ({\it Exploring a Web of Gravitational Theories through Gauge-Theory Methods}), as well as the Ragnar S\"{o}derberg Foundation (Swedish Foundations’ Starting Grant).


\appendix{A}{Gamma matrix conventions}

\seclab\appendixa

\subsec 10D gamma matrices

\subseclab \secbone

\noindent
We define the $16 \times 16$ gamma matrices,
\eqnn \gammaone
\eqnn \gammatwo
\eqnn \gammathree
$$
\eqalignno{
\gamma_{\alpha\beta}^0 &= 
\left(\matrix{
    -1 & 0 \cr
    0 & -1 
}\right)
\ ,& \gammaone \cr
\gamma_{\alpha\beta}^9 &=
\left(\matrix{
    1 & 0 \cr
    0 & -1 
}\right) 
\ ,&\gammatwo \cr
\gamma_{\alpha\beta}^{i=1,\ldots,8} &=
\left(\matrix{
    0 & \sigma^i \cr
    (\sigma^i)^T & 0
}\right) 
\ ,&\gammathree \cr
}
$$
where the superscript $T$ refers to transposition, and the $\sigma$ matrices are defined by,
\eqnn \sigmaone
\eqnn \sigmatwo
\eqnn \sigmathree
\eqnn \sigmafour
\eqnn \sigmafive
\eqnn \sigmasix
\eqnn \sigmaseven
\eqnn \sigmaeight
$$
\eqalignno{
\sigma^1 &=  \tau^2\otimes\tau^2\otimes\tau^2 \ ,& \sigmaone \cr
\sigma^2 &=  1\otimes\tau^1\otimes\tau^2 \ ,& \sigmatwo \cr
\sigma^3 &=  1\otimes\tau^3\otimes\tau^2 \ ,& \sigmathree \cr
\sigma^4 &=  \tau^1\otimes\tau^2\otimes1 \ ,& \sigmafour \cr
\sigma^5 &=  \tau^3\otimes\tau^2\otimes1 \ ,& \sigmafive \cr
\sigma^6 &=  \tau^2\otimes1\otimes\tau^1 \ ,& \sigmasix \cr
\sigma^7 &=  \tau^2\otimes1\otimes\tau^3 \ ,& \sigmaseven \cr
\sigma^8 &=  1\otimes1\otimes1 \ .& \sigmaeight \cr
}
$$
The $\tau$ matrices are partly rescaled Pauli matrices,
\eqnn \taumatrices
$$
\eqalignno{
\tau^1 &= 
\left(\matrix{
    0 & 1 \cr
    1 & 0 
}\right) \ , \ \ \ \ \ 
\tau^2 = 
\left(\matrix{
    0 & 1 \cr
    -1 & 0 
}\right) \ , \ \ \ \ \
\tau^3 = 
\left(\matrix{
    1 & 0 \cr
    0 & -1 
}\right)
\ .  & \taumatrices 
}
$$

\subsec 11D gamma matrices

\subseclab \secbtwo
We abuse notation by referring to both the 11D and 10D gamma matrices with $\gamma$. In 11D we make the initial definition of the $32\times 32$ gamma matrices,
\eqnn \egammaone
\eqnn \egammatwo
$$
\eqalignno{
\gamma_{\alpha\beta}^{10} &= 
\left(\matrix{
    0 & i \cr
    i & 0 
}\right)
\ ,& \egammaone \cr
\gamma_{\alpha\beta}^{i=0,\ldots, 9} &=
\left(\matrix{
    i\gamma^i & 0 \cr
    0 & -i\gamma^i 
}\right) 
\ .&\egammatwo 
}
$$
Next we redefine $\gamma^9\leftrightarrow\gamma^{10}$ in order to have the light-cone directions defined as in eqn. \lcdirections. Additionally in order to have the block form for $\gamma^\pm$, we rotate all gamma matrices by $\gamma\to R\cdot\gamma\cdot R^T$ where
\eqnn \rotate
$$
\eqalignno{
R &= 
\left(\matrix{
    1 & 0 &0 & 0 \cr
    0 & 0 & 0 & 1 \cr
    0 & 0 & 1 & 0 \cr
    0 & 1 & 0 & 0 
}\right)
\ .& \rotate 
}
$$
Finally in 11D we have a charge conjugation matrix,
\eqnn \charge
$$
\eqalignno{
C^{\alpha\beta} &= 
\left(\matrix{
    0 & 1  \cr
    -1 & 0 
}\right)
\ .& \charge 
}
$$
Our conventions for products of gamma matrices keep expressions as similar as possible to 10D gamma matrix products. So our definitions of forms are, for example, $(\gamma^{ab})_\alpha{}^{\beta} = (\gamma^{[a})_{\alpha\delta}(\gamma^{b]})^{\delta\beta}$. At times we have to raise and lower spinor indices using the charge conjugation matrix. We always raise or lower the right-most index in gamma matrix products, for example
\eqnn \indexraise
$$
\eqalignno{
(\gamma^{ab})^{\alpha\beta} &= 
C^{\beta\gamma}(\gamma^{ab})^{\alpha}{}_{\gamma}  = - (\gamma^{ab}C)^{\alpha\beta} \ ,& \indexraise 
}
$$
where in the last equality we used that the charge conjugation matrix is antisymmetric. Spinor products are written such that, if present, the charge conjugation is contracted in to the right-most spinor, so for instance
\eqnn \spinorprod
$$
\eqalignno{
(\lambda\gamma^{ab}\lambda) &= 
(\lambda\gamma^{ab}C\lambda) \ ,& \spinorprod 
}
$$
with ordinary matrix multiplication inside the parenthesis.

\appendix{B}{The Siegel vertex operator for 10D super-Yang-Mills}

\seclab\appendixa

\noindent In search for an alternative manifestly supersymmetric description of superstring theory free of the quantization problems presented by the Green-Schwarz superstring, Siegel proposed a completely first-order formulation for the particle-limit of the latter \siegelabcparticle. The worldline variables of this proposal consist of the coordinates $(X^m, \theta^{\alpha})$, and their respective conjugate momenta $(P^{m}, p_{\alpha})$, subject to the constraints
\eqnn \abcconstraints
$$
\eqalignno{
{\cal A} = P^{m}P_{m} \ \ , \ \ {\cal B} &= (\gamma^{m}d)_{\alpha}P_{m} \ \ , \ \ {\cal C}^{mnp} = (d\gamma^{mnp}d) \ ,& \abcconstraints
}
$$
where $d_{\alpha}$ is the familiar fermionic constraint of the Brink-Schwarz superparticle. Throughout this Appendix we will use letters from the middle/beginning of the Latin/Greek alphabet to denote $SO(1,9)$ vector/spinor indices. This new superparticle model was later shown to correctly reproduce the massless states of the open superstring when quantized in light-cone gauge \equivalencestrings. The super-Yang-Mills vertex operator was thus found to be described by
\eqnn \siegelgluon
$$
\eqalignno{
V &= P^{m}A_{m} + d_{\alpha}W^{\alpha} \ ,& \siegelgluon
}
$$
The objects $A_{m}$, $W^{\alpha}$ in \siegelgluon\ are the familiar 10D super-Yang-Mills superfields associated to the gluon and gluino states, respectively. They satisfy the superspace equations of motion
\eqnn \tendsym
$$ \eqalignno{
D_{\alpha}A_{\beta} + D_{\beta}A_{\alpha} = (\gamma^{m})_{\alpha\beta}A_{m} \ &, \ \ \ \ D_{\alpha}A_{m} = \partial_{m}A_{\alpha} + (\gamma_{m}W)_{\alpha}  \ ,\cr
D_{\alpha}W^{\beta} = -{1\over 4}(\gamma^{mn})_{\alpha}{}^{\beta}F_{mn} \ &, \ \ \ \ D_{\alpha}F_{mn} = 2(\gamma_{[m}\partial_{n]}W)_{\alpha}  \ ,& \tendsym
}
$$
where $A_{\alpha}$ is the lowest-dimensional component of the super-gauge connection, and $F_{mn}$ is the field-strength superfield. After fixing the Harnad-Shnider gauge $\theta^{\alpha}A_{\alpha} = 0$, these equations provide a solvable system of recursive relations which yield the superspace expansion coefficients of all the 10D super-Yang-Mills superfields, at all order in $\theta$ \refs{\tsimpistend,\mafrathesis}.

\medskip
\noindent Next we study the vertex operator \siegelgluon\ in light-cone gauge, and show it matches the light-cone gauge operators of the Brink-Schwarz superparticle.

\subsec Light-cone gauge

\subseclab \secapp

\noindent The light-cone gauge conditions on the worldline fields read
\eqnn \aplightconeeqnone
\eqnn \aplightconeeqntwo
$$
\eqalignno{
X^{+} &= x_{0}^{+} + P^{+}\tau \ ,& \aplightconeeqnone\cr
(\gamma^{+}\theta)_{\alpha} &= 0 \ ,& \aplightconeeqntwo
}
$$
Using the $SO(8)$ splitting $\theta^{\alpha} = (\theta^{a}, \bar{\theta}^{\dot{a}})$, where $a$, $\dot{a}$ are respectively $SO(8)$ chiral and antichiral spinor indices, one can write eqn. \aplightconeeqntwo\ in the equivalent form $\bar{\theta}^{\dot{a}} = 0$. This is easily seen to be the case in the basis where the gamma matrices are represented as in appendix A.
\medskip
\noindent The supersymmetric derivative and charge are denoted by $d_{\alpha}$ and $q_{\alpha}$ respectively, and defined as
\eqnn \ddefinition
\eqnn \qdefinition
$$
\eqalignno{
d_{\alpha} &= p_{\alpha} + {1 \over 2}(\gamma^{m}\theta)_{\alpha}P_{m} \ ,& \ddefinition\cr
q_{\alpha} &= p_{\alpha} - {1\over 2}(\gamma^{m}\theta)_{\alpha}P_{m} \ ,& \qdefinition
}
$$
Their forms in light-cone gauge are given by
\eqnn \dlightcone
\eqnn \qlightcone
$$
\eqalignno{
\bar{d}_{\dot{a}} = \bar{p}_{\dot{a}} + {1\over 2}(\sigma^{i})_{\dot{a}a}\theta^{a}P_{i} \ \  &, \ \ d_{a} = p_{a} - {\sqrt{2}\over 2}\theta_{a}P^{+} \ ,& \dlightcone\cr
\bar{q}_{\dot{a}} = \bar{p}_{\dot{a}} - {1\over 2}(\sigma^{i})_{\dot{a}a}\theta^{a}P_{i}\ \  &,  \ \ q_{a} = p_{a} + {\sqrt{2}\over 2}\theta_{a}P^{+} \ ,& \qlightcone
}
$$
where $d_{\alpha} = (d_{a},\bar{d}_{\dot{a}})$, $q_{\alpha} = ( q_{a},\bar{q}_{\dot{a}})$ and we use $i,j,k$ for $SO(8)$ vector indices. Likewise, the non-vanishing components of the ${\cal C}$-constraint in the light-cone frame read
\eqnn \ceqnone 
$$
\eqalignno{
{\cal C}^{+ij} = \sqrt{2}\,d_{a}(\sigma^{ij})^{ab}d_{b} \ \ &, \ \
{\cal C}^{ijk} = 2\,d_{a}(\sigma^{ijk})^{a\dot{a}}\bar{d}_{\dot{a}} \ ,& \ceqnone
}
$$
Moreover, the $SO(8)$ spinor variables $S^{a}$ of the light-cone gauge Brink-Schwarz superparticle, are related to the supersymmetric charge via
\eqnn \sandq
$$
\eqalignno{
S^{a} &= {q^{a}\over \sqrt{\sqrt{2}P^{+}}} \ ,& \sandq
}
$$
As usual, we will assume $k^{+} \rightarrow 0$. In order for the component $k^{-}$ to remain finite, one must have $k^{i}k^{i} \rightarrow 0$. This means that the momentum will be taken to be complex, and be restricted to take real values again in our final formulae. The use of this configuration and the gauge symmetry $\delta \epsilon_{m} = \partial_{m}\lambda$, allows one to set $\epsilon^{+} \rightarrow 0$. Similarly, the transversality condition requires $\epsilon^{i} k^{i} \rightarrow 0$, and thus $\epsilon^{-}$ is finite. Analogously, the equation of motion of the gluino imposes that its $SO(8)$ components are related to each other. All in all, one has
\eqnn \lightconekin
$$
\eqalignno{
k^{+} = 0 \ , \ \  k^{-} = {k^{i}k^{i}\over2k^{+}} \ &, \ \ \epsilon^{+} = 0 \ , \ \ \epsilon^{-} = {\epsilon^{i}k^{i}\over k^{+}} \ , \ \ \chi_{a} = -{1 \over \sqrt{2}}(\sigma^{i})_{a\dot{a}}k_{i}\bar{\xi}^{\dot{a}} \ , \ \  \bar{\xi}^{\dot{a}} = {\bar{\chi}^{\dot{a}}\over k^{+}} \ , & \cr 
& & \lightconekin
}
$$
where $\chi^{\alpha} = (\chi^{a},\bar{\chi}^{\dot{a}})$ is the gluino field.


\medskip
Let us now analyze the covariant vertex operator \siegelgluon\ in light-cone gauge. To this end, we first list the $\theta$-expansions of the superfields $A_{m}$, $W^{\alpha}$ (see \refs{\tsimpistend,\mafrathesis}),
\eqnn \amsuperfield
\eqnn \walphasuperfield
$$
\eqalignno{
A_{m} &= \epsilon_{m} - (\chi\gamma_{m}\theta) - {1\over 8}(\theta\gamma_{m}\gamma^{pq}\theta)f_{pq} + {1\over 12}(\theta\gamma_{m}\gamma^{pq}\theta)(\partial_{p}\chi\gamma_{q}\theta) & \cr 
& + {1 \over 192}(\theta\gamma_{mrs}\theta)(\theta\gamma^{spq}\theta)\partial^{r}f_{pq} + O(\theta^5) \ ,& \amsuperfield \cr
W^{\alpha} &= \chi^{\alpha} - {1\over 4}(\gamma^{mn}\theta)^{\alpha}f_{mn} + {1\over 4}(\gamma^{mn}\theta)^{\alpha}(\partial_{m}\chi\gamma_{n}\theta) + {1\over 48}(\gamma^{mn}\theta)^{\alpha}(\theta\gamma_{n}\gamma^{pq}\theta)\partial_{m}f_{pq} & \cr 
& -{1\over 96}(\gamma^{mn}\theta)^{\alpha}(\theta\gamma_{n}\gamma^{pq}\theta)(\partial_{m}\partial_{p}\chi\gamma_{q}\theta) + O(\theta^5) \ .& \walphasuperfield
}
$$
It is not hard to see that $A^{+} = 0$ in light-cone gauge. Therefore, the vertex operator \siegelgluon\ can be written as
\eqnn \covvertexlightconetwo
$$
\eqalignno{
V &= -P^{+}A^{-} + P^{i}A^{i} + d_{a}W^{a} + \bar{d}_{\dot{a}}\bar{W}^{\dot{a}} \ . & \covvertexlightconetwo
}
$$
For simplicity, let us focus on the bosonic sector. The fermionic counterpart directly follows from supersymmetry. Using eqns. \amsuperfield, \walphasuperfield, one finds that all the terms of the expansions vanish except for those linear and quadratic in $\theta^{\alpha}$. Explicitly,
\eqnn \expanding
$$
\eqalignno{
V &= -P^{+}\bigg[\epsilon^{-} + {\sqrt{2}\over 4}(\theta\sigma^{ij}\theta)k_{i}\epsilon_{j}\bigg] + P^{i}\epsilon^{i} + d_{a}\bigg[- {1\over 2}(\sigma^{ij}\theta)^{a}k_{i}\epsilon_{j}\bigg] & \cr
&= -P^{+}\epsilon^{-} + P^{i}\epsilon^{i} - {\sqrt{2}\over 4}(\theta\sigma^{ij}\theta)k_{i}\epsilon_{j}P^{+} - {1\over 2}(d\sigma^{ij}\theta)k_{i}\epsilon_{j} \ .& \expanding
}
$$
The relations \dlightcone, \qlightcone\ define $\theta^{a}$ by the simple formula
\eqnn \thetawithdq
$$
\eqalignno{
\theta^{a} &= {1\over \sqrt{2}P^{+}}(q^{a} - d^{a}) \ .& \thetawithdq
}
$$
After plugging \thetawithdq\ into \expanding, and using \ceqnone, one is left with
\eqnn \simplifying
$$
\eqalignno{
V &= -P^{+}\epsilon^{-} + P^{i}\epsilon^{i} - {\sqrt{2}\over 8P^{+}}(q\sigma^{ij}q)k_{i}\epsilon_{j} + {\sqrt{2}\over 4P^{+}}(q\sigma^{ij}d)k_{i}\epsilon_{j} - {1\over 2\sqrt{2}P^{+}}(d\sigma^{ij}q)k_{i}\epsilon_{j} & \cr
&= -P^{+}\epsilon^{-} + P^{i}\epsilon^{i} - {\sqrt{2}\over 8P^{+}}(q\sigma^{ij}q)k_{i}\epsilon_{j} \ . & 
}
$$
Finally, the use of eqn. \sandq\ allows one to conclude that
\eqnn \finalightcone
$$ \eqalignno{
V &= -P^{+}\epsilon^{-} + P^{i}\epsilon^{i} - {1\over 4}(S\sigma^{ij}S)k_{i}\epsilon_{j}  \ , & \simplifying
}
$$
which is exactly the gluon vertex operator in the light-cone gauge Brink-Schwarz worldline framework.

\listrefs

\bye